\documentclass[journal,12pt,draftclsnofoot,onecolumn]{IEEEtran}

\usepackage[noadjust]{cite}

\usepackage{graphicx}
\usepackage{tabularx}
\usepackage{tablefootnote}
\usepackage{enumitem}
\usepackage[T1]{fontenc}
\usepackage{adjustbox}
\usepackage{booktabs, multirow}
\usepackage[skip=1ex]{caption}
\usepackage{enumitem}
\usepackage[font=normalsize]{caption}
\usepackage{amsmath}

\usepackage[usestackEOL]{stackengine}
\usepackage{amssymb}
\usepackage[scaled=0.9]{DejaVuSansMono}
\usepackage{listings}
\usepackage{amsmath}
\usepackage{amsfonts}

\usepackage{amsthm}

\usepackage{bm}
\usepackage{url}
\usepackage{mathtools}
\usepackage{balance}
\allowdisplaybreaks
\usepackage{algorithm}
\usepackage[noend]{algpseudocode}
\usepackage{etoolbox}
\usepackage[mathscr]{euscript}
\usepackage{calc}
\usepackage{pgfplots}
\usepackage{tikz}
\usepackage{setspace}
\usepackage{pgfplotstable}
\DeclareMathAlphabet{\pazocal}{OMS}{zplm}{m}{n}
\DeclareSymbolFont{missing}{OML}{cmr}{m}{n}
\DeclareMathSymbol{\ell}{\mathord}{missing}{'140}

\usetikzlibrary{calc}
\usetikzlibrary{patterns}
\usetikzlibrary{spy,backgrounds}
\pgfplotsset{grid style={dotted,gray}}
\DeclareMathOperator*{\argmin}{arg\,min}
\DeclareMathOperator*{\argmax}{arg\,max}
\newcommand{\X}{\mathscr}

\newcommand{\maximize}{%
  \mathopen{}\operatorname*{maximize}%
}

\newcommand{\subjto}{\textup{subject to}}

\newcounter{problem}
\newcounter{save@equation}
\newcounter{save@problem}

\newtheorem{lemma}{Lemma}
\newtheorem{theorem}{Theorem}

\newtheorem{definition}{Definition}

\newtheorem*{remark}{Remark}
\newlength{\depthofsumsign}
\newlength{\totalheightofsumsign}
\newlength{\heightanddepthofargument}

\makeatletter

\tikzset{reset label anchor/.code={%
    \let\tikz@auto@anchor=\pgfutil@empty
    \def\tikz@anchor{#1}
  },
  reset label anchor/.default=center
}

\let\save@mathaccent\mathaccent
\newcommand*\if@single[3]{%
  \setbox0\hbox{${\mathaccent"0362{#1}}^H$}%
  \setbox2\hbox{${\mathaccent"0362{\kern0pt#1}}^H$}%
  \ifdim\ht0=\ht2 #3\else #2\fi
}
\newcommand*\rel@kern[1]{\kern#1\dimexpr\macc@kerna}
\newcommand*\widebar[1]{\@ifnextchar^{{\wide@bar{#1}{0}}}{\wide@bar{#1}{1}}}
\newcommand*\wide@bar[2]{\if@single{#1}{\wide@bar@{#1}{#2}{1}}{\wide@bar@{#1}{#2}{2}}}
\newcommand*\wide@bar@[3]{%
  \begingroup
  \def\mathaccent##1##2{%
    \let\mathaccent\save@mathaccent
    \if#32 \let\macc@nucleus\first@char \fi
    \setbox\z@\hbox{$\macc@style{\macc@nucleus}_{}$}%
    \setbox\tw@\hbox{$\macc@style{\macc@nucleus}{}_{}$}%
    \dimen@\wd\tw@
    \advance\dimen@-\wd\z@
    \divide\dimen@ 3
    \@tempdima\wd\tw@
    \advance\@tempdima-\scriptspace
    \divide\@tempdima 10
    \advance\dimen@-\@tempdima
    \ifdim\dimen@>\z@ \dimen@0pt\fi
    \rel@kern{0.6}\kern-\dimen@
    \if#31
    \overline{\rel@kern{-0.6}\kern\dimen@\macc@nucleus\rel@kern{0.4}\kern\dimen@}%
    \advance\dimen@0.4\dimexpr\macc@kerna
    \let\final@kern#2%
    \ifdim\dimen@<\z@ \let\final@kern1\fi
    \if\final@kern1 \kern-\dimen@\fi
    \else
    \overline{\rel@kern{-0.6}\kern\dimen@#1}%
    \fi
  }%
  \macc@depth\@ne
  \let\math@bgroup\@empty \let\math@egroup\macc@set@skewchar
  \mathsurround\z@ \frozen@everymath{\mathgroup\macc@group\relax}%
  \macc@set@skewchar\relax
  \let\mathaccentV\macc@nested@a
  \if#31
  \macc@nested@a\relax111{#1}%
  \else
  \def\gobble@till@marker##1\endmarker{}%
  \futurelet\first@char\gobble@till@marker#1\endmarker
  \ifcat\noexpand\first@char A\else
  \def\first@char{}%
  \fi
  \macc@nested@a\relax111{\first@char}%
  \fi
  \endgroup
}

\newenvironment{problem}
{\setcounter{problem}{\value{save@problem}}%
  \setcounter{save@equation}{\value{equation}}%
  \let\c@equation\c@problem
  \subequations
}
{\endsubequations
  \setcounter{save@problem}{\value{equation}}%
  \setcounter{equation}{\value{save@equation}}%
}

\algnewcommand{\LineComment}[1]{\Statex \hskip\ALG@thistlm
  \(\triangleright\) #1}
\def\BState{\State\hskip-\ALG@thistlm}

\algdef{SE}[DOWHILE]{Do}{EndDoWhile}{\algorithmicdo}[1]{\algorithmicwhile\ #1}%
 \newcommand*{\algrule}[1][\algorithmicindent]{%
 \makebox[#1][l]{%
 \hspace*{.2em}
 \vrule height .75\baselineskip depth .3\baselineskip
 }
 }

   \newcount\ALG@printindent@tempcnta
   \def\ALG@printindent{%
   \ifnum \theALG@nested>0
   \ifx\ALG@text\ALG@x@notext
   \else
   \unskip
   \ALG@printindent@tempcnta=1
   \loop
   \algrule[\csname ALG@ind@\the\ALG@printindent@tempcnta\endcsname]%
   \advance \ALG@printindent@tempcnta 1
   \ifnum \ALG@printindent@tempcnta<\numexpr\theALG@nested+1\relax
   \repeat
   \fi
   \fi
 }
   \patchcmd{\ALG@doentity}{\noindent\hskip\ALG@tlm}{\ALG@printindent}{}{\errmessage{failed
   to patch}}
\patchcmd{\ALG@doentity}{\item[]\nointerlineskip}{}{}{}
   \tikzset{nomorepostaction/.code=\let\tikz@postactions\pgfutil@empty}

   \long\def\ifnodedefined#1#2#3{%
   \@ifundefined{pgf@sh@ns@#1}{#3}{#2}%
 }
\usetikzlibrary{snakes}
\tikzstyle{printersafe}=[snake=snake,segment amplitude=0 pt]
\usetikzlibrary{decorations.pathmorphing}
\tikzstyle{printersafe}=[decoration={snake,amplitude=0pt}]
\pgfplotsset{
  discontinuous/.style={
    scatter,
    scatter/@pre marker code/.code={
      \ifnodedefined{marker}{
        \pgfpointdiff{\pgfpointanchor{marker}{center}}%
        {\pgfpoint{0}{0}}%
        \ifdim\pgf@y<0pt
        \draw plot [mark=halfsquare right*] coordinates {(marker-|0,0)};
        \draw [densely dashed] (marker-|0,0) -- (0,0);
        \tikzset{options/.style={mark=*, fill=white}}
        \else
        \tikzset{options/.style={mark=*}}
        \fi
      }{
        \tikzset{options/.style={mark=*, fill=white}}        
      }
      \coordinate (marker) at (0,0);
      \begin{scope}[options]
      },
      scatter/@post marker code/.code={\end{scope}}
  }
}

\pgfplotsset{
  discontinuous1/.style={
    scatter,
    scatter/@pre marker code/.code={
      \ifnodedefined{marker}{
        \pgfpointdiff{\pgfpointanchor{marker}{center}}%
        {\pgfpoint{0}{0}}%
        \ifdim\pgf@y<0pt
        \draw plot [mark=halfcircle*] coordinates {(marker-|0,0)};
        \draw [densely dashed] (marker-|0,0) -- (0,0);
        \tikzset{options/.style={mark=*, fill=white}}
        \else
        \tikzset{options/.style={mark=*}}
        \fi
      }{
        \tikzset{options/.style={mark=*, fill=white}}        
      }
      \coordinate (marker) at (0,0);
      \begin{scope}[options]
      },
      scatter/@post marker code/.code={\end{scope}}
  }
}

\makeatother

\usetikzlibrary{automata}
\usetikzlibrary{calc,fit,shapes.geometric}
\pgfdeclarelayer{signal}
\pgfsetlayers{signal,main}
\usetikzlibrary{snakes}
\tikzstyle{printersafe}=[snake=snake,segment amplitude=0 pt]
\usetikzlibrary{arrows}
\newcounter{cntr}
\usetikzlibrary{calc}
\usetikzlibrary{decorations.pathreplacing,decorations.markings,shapes.geometric}
\tikzset{naming/.style={align=center}}
\tikzset{antenna/.style={insert path={-- coordinate (ant#1) ++(0,0.5) -- +(135:0.5) + (0,0) -- +(45:0.5)}}}
\tikzset{station/.style={naming,draw,shape=dart,shape border rotate=90, minimum width=20mm, minimum height=20mm,outer sep=0pt,inner
    sep=3pt}}
\tikzset{mobile/.style={naming,draw,shape=rectangle,minimum width=12mm,minimum height=6mm, outer sep=0pt,inner sep=3pt}}
\tikzset{radiation/.style={{decorate,decoration={expanding waves,angle=90,segment length=6pt}}}}

\tikzset{
  every pin/.style={rectangle,rounded corners=3pt,font=\footnotesize},
  small dot/.style={fill=black,circle,scale=0.5}
}

\tikzset{
  invisible/.style={opacity=0},
  visible on/.style={alt={#1{}{invisible}}},
  alt/.code args={<#1>#2#3}{%
    \alt<#1>{\pgfkeysalso{#2}}{\pgfkeysalso{#3}} 
  },
}

\tikzset{pics/.cd,
  SBS/.style={code={
      \begin{scope}[local bounding box=#1]
        \fill [pic actions/.try] (-1,0) -- (-1/2,3) -- (1/2, 3) -- (1,0) -- cycle;
        \fill [pic actions/.try] (-1/16,2) rectangle (1/16,4);
        \fill [pic actions/.try] (0,4) circle [radius=1/4];
        \foreach \i in {-1,1}
        \fill [shift=(90:4), xscale=\i]
        \foreach \r in {1,3/2,2}{
          (-45:\r) arc (-45:45:\r) -- (45:\r-1/10)
          arc(45:-45:\r-1/10) -- cycle
        };
      \end{scope}
    }},
  MBS/.style={code={
      \begin{scope}[local bounding box=#1]
        \fill [pic actions/.try] (-1,0) -- (-1/2,3) -- (1/2, 3) -- (1,0) -- cycle;
        \fill [pic actions/.try] (-1/16,2) rectangle (1/16,4);
        \fill [pic actions/.try] (0,4) circle [radius=1/4];
        \foreach \i in {-1,1}
        \fill [shift=(90:4), xscale=\i]
        \foreach \r in {1,3/2,2}{
          (-45:\r) arc (-45:45:\r) -- (45:\r-1/10)
          arc(45:-45:\r-1/10) -- cycle
        };
      \end{scope}
    }},
  SU/.style={code={
      \begin{scope}[local bounding box=#1]
        \fill [even odd rule, pic actions/.try]
        (-1,-5/2) -- (-1,-1/8) -- (1,-1/8) -- (1,-5/2)
        arc (360:180:1 and 1/4) -- cycle (-1,5/2) -- (-1,1/8) -- (1,1/8) -- (1,5/2)
        arc (0:180:1 and 1/4) -- cycle (-3/4, 9/4) -- (-3/4, 3/8) -- (3/4, 3/8) -- (3/4, 9/4)
        arc (0:180:3/4 and 1/8)-- cycle
        \foreach \i in {-1,0,1}{\foreach \j in {1,2,3}{
            (-\i*1/2-3/16,-\j/2-3/4) rectangle ++(3/8, 3/8)
          }
        }
        (-1/2,-3/4) rectangle (1/2, -1/4);
      \end{scope}
    }},
  MU/.style={code={
      \begin{scope}[local bounding box=#1]
        \fill [even odd rule, pic actions/.try]
        (-1,-5/2) -- (-1,-1/8) -- (1,-1/8) -- (1,-5/2)
        arc (360:180:1 and 1/4) -- cycle (-1,5/2) -- (-1,1/8) -- (1,1/8) -- (1,5/2)
        arc (0:180:1 and 1/4) -- cycle (-3/4, 9/4) -- (-3/4, 3/8) -- (3/4, 3/8) -- (3/4, 9/4) arc (0:180:3/4 and 1/8)-- cycle
        \foreach \i in {-1,0,1}{
          \foreach \j in {1,2,3}{
            (-\i*1/2-3/16,-\j/2-3/4) rectangle ++(3/8, 3/8)
          }
        }
        (-1/2,-3/4) rectangle (1/2, -1/4);
      \end{scope}
    }},
  SIGNAL/.style={code={
      \begin{scope}[local bounding box=#1]
        \fill [pic actions/.try]
        (0,-3) -- (-1,1/2) -- (1/8,1/4) -- (0,3) -- (1,-1/2) -- (-1/8,-1/4) -- cycle;
      \end{scope}
    }},
  queuei/.style={code={
      \begin{scope}
        \stepcounter{cntr}
        \node[inner sep=0pt, outer sep=0pt,draw,rectangle split,rectangle split horizontal,minimum height=0.5cm,rectangle split parts=3]
        (queue-\thecntr) [pic actions] {};
        \draw
        (queue-\thecntr.north west) -- ++(-0.2cm,0)
        (queue-\thecntr.south west) -- ++(-0.2cm,0);
        \node[above] at ([xshift=-0.5cm]queue-\thecntr.north)
        {$Q_#1$};
      \end{scope}
    }}
}
\colorlet{sky blue}{blue!60!cyan!75!black}
\colorlet{dark blue}{blue!50!cyan}
\colorlet{chameleon}{olive!75!green}
\tikzset{signal/.style={->,draw=black, line width=0.05em, dashed,printersafe}}

\newsavebox{\mybox}
\begin{document}

\title{Resource Allocation in Green Dense Cellular Networks: Complexity and Algorithms}

\author{\IEEEauthorblockN{Zoubeir Mlika, Elmahdi Driouch and Wessam Ajib}}

\maketitle

\begin{abstract}
  This paper studies the problem of user association, scheduling and channel allocation in dense cellular networks with energy harvesting base stations (EBSs). In this problem, the EBSs are powered solely by renewable energy and each user has a request for downloading data of certain size with a deadline constraint. The objective is to maximize the number of associated and scheduled users while allocating the available channels to the users and respecting the energy and deadline constraints. First, the computational complexity of this problem is characterized by studying its $\X{NP}$-hardness in different cases. Next, efficient algorithms are proposed in each case. The case of a single channel and a single EBS is solved using two polynomial-time optimal algorithms---one for arbitrary deadlines and a less-complex one for common deadlines. The case of a single channel and multiple EBSs is solved by proposing an efficient constant-factor approximation algorithm. The case of multiple channels is efficiently solved using a heuristic algorithm. Finally, our theoretical analysis is supplemented by simulation results to illustrate the performance of the proposed algorithms.
\end{abstract}

\begin{IEEEkeywords}
  User association, scheduling, channel allocation, energy harvesting, approximation algorithms, $\X{NP}$-hardness.
\end{IEEEkeywords}

\IEEEpeerreviewmaketitle


\section{Introduction}\label{section:introduction}
High spectral efficiency and ultra-low latency are key requirements of 5th generation (5G) cellular networks~\cite{Andrews:32:6,Damnjanovic:18:3}. Initial 5G deployments will focus on enhanced mobile broadband (eMBB) applications with the spectral efficiency being one of the most important key performance indicators~\cite{Liu:35:6}. Dense cellular networks (DCNs), where base stations (BSs) are densely deployed in a small geographic area, are considered as an ideal solution to reach such high spectral efficiency. In DCNs, frequency channels are generally allocated based on two approaches: a full frequency reuse and a fractional frequency reuse. In the full reuse approach, all BSs operate on the same channel, which may result in spectral efficiency improvements if the interference is carefully managed~\cite{Elsawy:28:36,Zhuang:2931:2942}. Additionally, the BSs consume an important amount of energy~\cite{Wang:11:2}. Consequently, an efficient resource allocation for interference management and reduced energy consumption are of extreme importance in DCNs. In this paper, resource allocation in DCNs refers to channel allocation, user association, and scheduling, which are three coupled problems that are very hard to solve jointly. Further, energy harvesting and self-powered BSs are deployed in DCNs in order to maximize the use of green energy. 

In this paper, we focus on the Resource Allocation problem with Energy and Deadline constraints (we name it \textit{RAED}). In \textit{RAED}, each user requests to download some data of a given size before a hard deadline. The objective of \textit{RAED} is to associate and schedule as many users as possible while allocating the available channels to the users subject to the constraints imposed by the request deadlines and the EBSs available energy levels. Due to uncertain and limited levels of the harvested energy and to the coupled nature of this problem, it is very challenging to solve \textit{RAED} in DCNs. To the best of our knowledge, previous research literature did not deal with such problem under the same objective and constraints as considered in this paper. 

\subsection{Related Work}\label{ssection:literature}
The following summarizes the most important work related to our research. In~\cite{Zhuang:2931:2942} the authors propose a scalable resource allocation approach to solve the channel allocation and user association problem in heterogeneous networks (HetNets) with the objective of minimizing the average packet delay. Their approach iteratively solves a convex optimization problem and an hyper-graph coloring problem. Similar problems to~\cite{Zhuang:2931:2942} are studied in~\cite{Zhuang:5470:5483,Zhuang:34:4}. In~\cite{Zhao:5825:5837}, the authors study the problem of channel allocation and power control in non-orthogonal multiple access networks. They use a matching game to design a two-sided exchange-stable algorithm to solve the channel allocation problem. Also, they use sequential convex programming to solve the problem of power control. The authors of~\cite{Lin:1025:1039} study the user association problem where they assume two scenarios: full channel reuse and fractional (orthogonal) channel reuse. They formulate a network utility maximization problem and they use stochastic geometry to obtain the analytical user association bias factors and the channel partition ratios. The works in~\cite{Zhuang:2931:2942,Zhuang:5470:5483,Zhuang:34:4,Zhao:5825:5837,Lin:1025:1039} solve the channel allocation or the user association problem without considering energy harvesting BSs nor the scheduling problem. In~\cite{Huang:1235:1249}, the authors study multicast scheduling in cellular networks under deadline constraints. Packet scheduling with common deadline is investigated in~\cite{Deshmukh:3661:3674} with the objective of energy minimization. Both~\cite{Huang:1235:1249} and~\cite{Deshmukh:3661:3674} deal exclusively with scheduling without any reference to user association. In~\cite{Krishnasamy:307:314}, channel allocation and scheduling is considered in device-to-device (D2D)-enabled DCNs where cellular users have common delay requirements. In~\cite{Wang:6:5}, the authors consider the problem of real-time packet scheduling in long term evolution advanced (LTE-A) networks. The proposed scheduling algorithm is based on the almost blank subframe (ABS) method to manage the interference. In~\cite{Sciancalepore:193:201}, the authors study the scheduling problem in LTE networks based also on ABS method. They propose a semi-distributed algorithm to achieve low overhead. In~\cite{Mlika:21:12}, a problem similar to \textit{RAED} but without channel allocation nor EBSs is considered. The authors show that considering different arrival times of users requests renders the problem $\X{NP}$-hard even for a single BS network. Next, they develop a constant-factor approximation algorithm to solve the problem. Other related work include packet or job scheduling in the context of scheduling theory~\cite{Chetto:2:2,Wang:4:3}. For example, \cite{Chetto:2:2} study the problem of real-time job scheduling in an energy harvesting system. All mentioned previous works assume different models and objectives from the model presented in this paper.

Note that most previous research works do not provide theoretical algorithmic analysis of the resource allocation problem in DCNs, e.g., neither $\X{NP}$-hardness nor approximation algorithms were proposed. In this research, we fill this gap (i) by analyzing the computational complexity of \textit{RAED} in different practical cases and (ii) by proposing efficient algorithms with worst-case performance guarantees.

\subsection{Contributions}\label{sec:contr}
This work studies \textit{RAED} and its computational complexity in different cases; depending on the number of channels and/or the number of EBSs. 
The main contributions of this work are summarized in the following list.
\begin{itemize}
	\item We model \textit{RAED} as an integer linear program (ILP) and characterize its computational complexity by studying its $\X{NP}$-hardness considering the following four cases: (i) the single channel and single EBS (SCSB) case, (ii) the single channel and multiple EBSs (SCMB) case, (iii) the multiple channels and single EBS (MCSB) case, and (iv) the multiple channels and multiple EBSs (MCMB) case.
	\item For the case of SCSB, we propose
\begin{itemize}
	\item an optimal polynomial-time (polytime) algorithm for arbitrary deadlines; and
	\item a less complex optimal polytime algorithm for common deadlines.
\end{itemize}
	\item For the case of SCMB, we propose an efficient constant-factor approximation algorithm.
	\item For the case of MCSB and MCMB, we propose an efficient heuristic algorithm. 
	\item Finally, we show that a preemptive scheduling solution to \textit{RAED} can be modified, in polytime, to obtain a non-preemptive one.
\end{itemize}

\subsection{Organization}\label{sec:org}
The paper is organized as follows. Section~\ref{sec:model} presents the system model and introduces \textit{RAED}. Section~\ref{sec:SAED:pbform} formulates \textit{RAED} and characterizes its computational complexity. Section~\ref{sec:SAED:BS} studies \textit{RAED} in the case of single channel and proposes optimal and approximation algorithms to solve it. Section~\ref{sec:RAED} studies \textit{RAED} in the case of multiple channels, proposes heuristic algorithms and discusses the non-preemptive scheduling scenario. Section~\ref{sec:sim} presents the simulations results that illustrate the performance of the proposed algorithms. Finally, Section~\ref{sec:cl} draws some conclusions.


\section{System Model}\label{sec:model}
We consider a dense cellular network (DCN) composed of $B$ single antenna energy harvesting base stations (EBSs) denoted by the set $\X{B}=\{1,2,\ldots,B\}$. The total bandwidth is divided into a set of $C$ orthogonal channels $\X{C}=\{1,2,\ldots,C\}$ that can be used by the EBSs for downlink transmission. The transmit power of EBS $b$ using any channel is fixed to $P_b$ similarly to~\cite{Zhuang:2931:2942}. Note that the single antenna model is worth studying for the following reasons: (i) it helps to understand the intrinsic difficulty of the problem in the multiple antenna model, (ii) it helps to characterize the structure of the solutions for the multiple antenna model, and (iii) it serves as guideline for the multiple antenna model, i.e. the proposed algorithms for the single antenna model will serve as benchmarks for the algorithms developed in the multiple antenna model.

There are $U$ users in the DCN denoted by the set $\X{U}=\{1,2,\ldots,U\}$. Time is divided into frames where each frame is composed of $T$ slots of duration $\tau$ seconds each. The optimization process to solve the resource allocation problem is performed at the start of each frame. Let $\X{T}=\{1,2,\ldots,T\}$ be the current frame. Every user $u\in\X{U}$ has a data request $(s_u,d_u)$ where $s_u$ is its size in bits and $d_u\in\X{T}$ is its deadline. The EBSs are self-powered thanks to their energy harvesting capabilities. At each time slot $t\in\X{T}$, the amount of harvested energy of EBS $b$ is denoted by $E_{b,t}$ that is stored in EBS's battery assumed to have a large capacity~\cite{Shan:33:3}. In the rest of the paper, we normalize $E_{b,t}$ as $A_{b,t}\coloneq E_{b,t}/P_b$. Here, $A_{b,t}$ represents the minimum number of slots that can be used by EBS $b$ to transmit with power $P_b$ from slot $t$ (at least $A_{b,t}$ slots are available for scheduling starting from $t$). By this normalization, energy and slots are treated equivalently in this paper, i.e. when we say energy is available at $t$, it implies that there is at least one slot that can be used from  time $s\leqslant t$. Note that, energy arrival $E_{b,t}$ (and hence the number of slots $A_{b,t}$) can be any arbitrary non-negative value and it is not assumed to follow any particular distribution, i.e. we consider the general non-stochastic case.
\begin{figure}[ht!]
	\centering
	\includegraphics[scale=0.65]{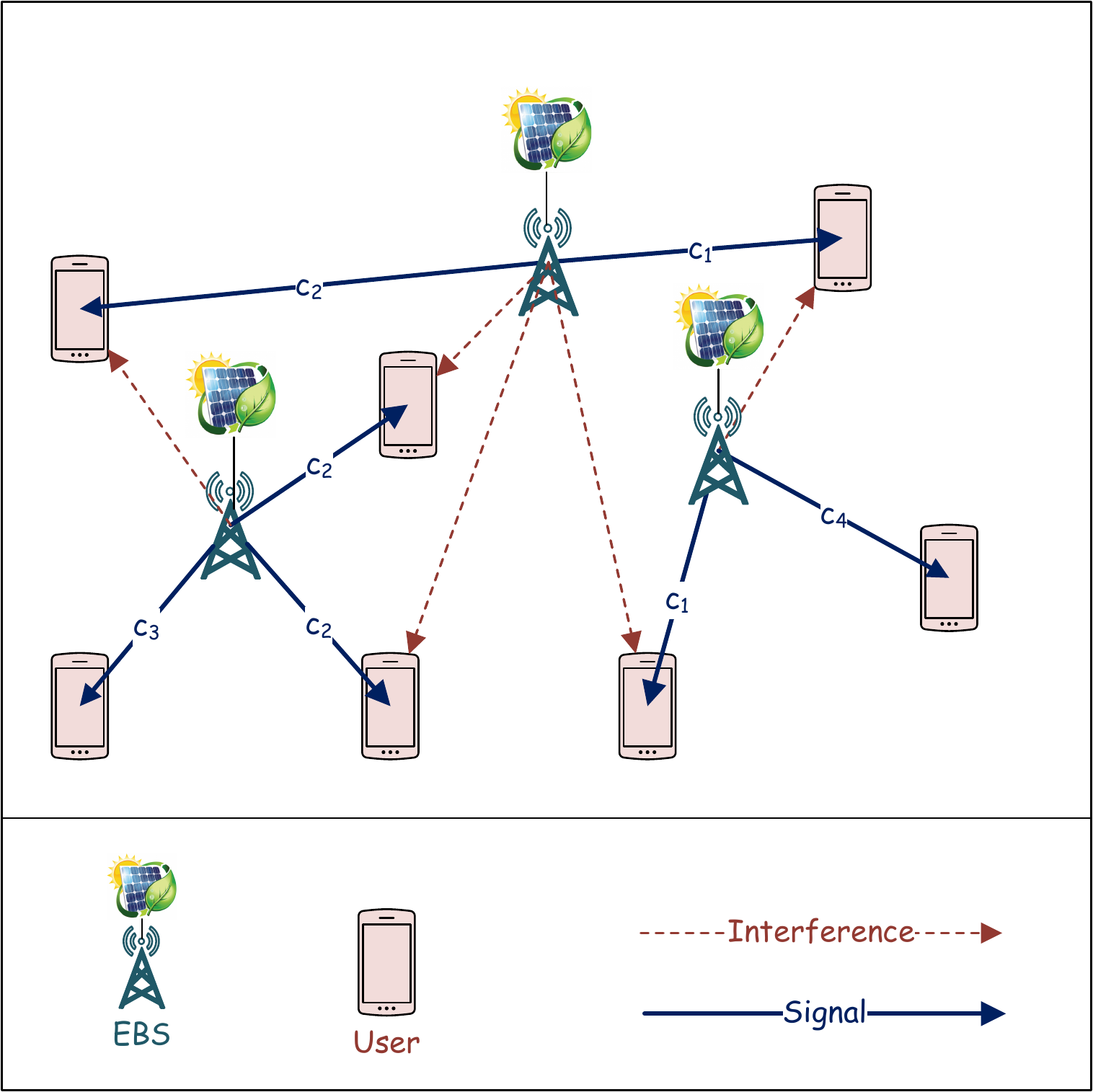}
	\caption{\footnotesize An example of the system model.}
	\label{model}
\end{figure}

When user $u$ is associated to EBS $b$ using channel $c$, the received signal to interference-plus-noise ratio (SINR) is given by~\cite{Zhuang:34:4}
\begin{align}
  \label{sinr:DL}
  \mathit{SINR}_{u,b}^c= \dfrac{P_bh_{u,b}^c}{\sigma^2+\sum_{b'\in\mathscr{B}\backslash\{b\}}P_{b'}h_{u,b'}^c},
\end{align}
where $\sigma^2$ is the power of the additive white Gaussian noise (AWGN) and $h_{u,b}^c$ is the channel power gain between user $u$ and EBS $b$ using channel $c$, which takes into account large-scale pathloss propagation effect on a slow timescale~\cite{Zhuang:34:4}. From~\eqref{sinr:DL}, the achievable data rate, $R_{u,b}^c$ (in bps/Hz) of user $u$ when associated to EBS $b$ using channel $c$ can be calculated as
\begin{align}\label{rate:DL}
  R_{u,b}^c=\log_2\left(1+\mathit{SINR}_{u,b}^c\right).
\end{align}

User $u$ fulfills its request, when associated to EBS $b$ using channel $c$, if it is allocated $\nu_{u,b}^c$ slots, where $\nu_{u,b}^c$ is defined as
\begin{align}\label{required:slots}
  \nu_{u,b}^c\coloneq\left\lceil\dfrac{s_uC}{\tau WR_{u,b}^c}\right\rceil,
\end{align}
where $W$ is the total bandwidth in Hz. Hence, in order for user $u$ to download its $s_u$ bits, it needs to be associated to EBS $b$ using channel $c$ and scheduled for $\nu_{u,b}^c$ slots before its deadline $d_u$ and whenever energy is available at EBS $b$. In the rest of the paper, when there is only one EBS (or one channel) in the network, we drop the subscript $b$ (or the superscript $c$). An example of the system model is given in Fig.~\ref{model}, where there are seven users, three EBSs and four frequency channels $\{c_1, c_2, c_3, c_4\}$. There is an EBS that schedules two users on channel $c_2$ at different time slots.

The objective of this work is to maximize the number of associated and scheduled users in the current frame $\X{T}$ while allocating the channels to the users and satisfying the users demands and respecting the deadlines and energy constraints. \textit{RAED} is solved in the \textit{perfectly-predictable energy} arrival model where future energy arrivals are perfectly predicted~\cite{Hentati}. Other inputs to \textit{RAED} are assumed to be known at the beginning of the frame. The non-stochastic \textit{online} scenario where all inputs are assumed to be unknown is left for future work. Note that the proposed offline algorithms can help us to develop efficient online competitive algorithms in future works. Throughout the rest of the paper, the term \textit{served} users is used instead of associated and scheduled users. 

In the next section, we formulate the problem as an integer linear program and we analyze its computational complexity in different cases.

\section{Problem Formulation and Complexity Analysis}\label{sec:SAED:pbform}
\subsection{Problem Formulation}
In this subsection, \textit{RAED} is formulated as an integer linear program (ILP) to help solving it optimally and efficiently using branch-and-bound algorithm, since brute-force-based approaches are generally inefficient especially when $U$, $B$, $C$, or $T$ are large.

To formulate \textit{RAED} as an ILP, we introduce the following decision binary variable.
\begin{align*}
x_{u,b,t}^c\coloneq
\begin{cases}
1, & \text{if $u$ is associated to $b$ using channel $c$ at slot $t$,} \\
0, & \text{otherwise.}
\end{cases}
\end{align*}
Also, let $z_{b,t}$ be the decision integer variable that represents the amount of accumulated number of slots for EBS $b$ at slot $t$.

The constraints of \textit{RAED} can be formulated as follows.
\begin{itemize}
	\item User $u$ cannot use more than one channel when associated to EBS $b$
	\begin{align}\label{cns:1}
	x_{u,b,t}^c+x_{u,b,t'}^{c'}\leqslant1,\quad\forall c\neq c', (u,b,t,t',c,c')\in\X{U}\times\X{B}\times\X{T}^2\times\X{C}^2.\tag{P1c}
	\end{align}
	
	\item The users that are associated to EBS $b$ using channel $c$ should not interfere at slot $t$. Two users associated to EBS $b$ using channel $c$ are said to interfere if they are scheduled at the same slot. These constraints are expressed as
	\begin{align}\label{cns:2}
		\sum_{u\in\X{U}}x_{u,b,t}^c\leqslant1,\quad\forall(b,c,t)\in\X{B}\times\X{C}\times\X{T}.\tag{P1d}
	\end{align}
	
	\item User $u$ is associated to only one EBS using channel $c$. We can express these constraints as
	\begin{align}\label{cns:3}
		x_{u,b,t}^c+x_{u,b',t'}^{c'}\leqslant1,\quad\forall b\neq b', (u,b,b',t,t',c,c')\in\X{U}\times\X{B}^2\times\X{T}^2\times\X{C}^2.\tag{P1e}
	\end{align}
	
	\item The constraints about the update of the number of slots are formulated as
	\begin{align}\label{cns:4}
	z_{b,t+1}=z_{b,t}+A_{b,t+1}-\sum_{u\in\X{U}}\sum_{c\in\X{C}}x_{u,b,t}^c,\quad\forall(b,t)\in\X{B}\times\{1,\ldots,T-1\},\tag{P1f}
	\end{align}
	and the initial conditions are
	\begin{align}\label{cns:4:1}
	z_{b,1}=A_{b,1},\quad\forall b\in\X{B}.\tag{P1g}
	\end{align}
	
	\item Each user $u$ is associated to EBS $b$ using channel $c$ only when a positive amount of energy is available at $t$. Hence:
	\begin{align}\label{cns:5}
	x_{u,b,t}^c\leqslant z_{b,t},\quad\forall(u,b,c,t)\in\X{U}\times\X{B}\times\X{C}\times\X{T},\tag{P1h}
	\end{align}
	
	\item Each user $u$ requires $\nu_{u,b}^c$ slots when associated to EBS $b$ using channel $c$. These constraints can be written as
	\begin{align}\label{cns:6}
	x_{u,b,t}^c\sum_{s\in\X{T}}x_{u,b,s}^c=\nu_{u,b}^cx_{u,b,t}^c,\quad\forall(u,b,c,t)\in\X{U}\times\X{B}\times\X{C}\times\X{T}.
	\end{align}
	Note that we multiply both sides of~\eqref{cns:6} by $x_{u,b,t}^c$ because these constraints are only active when user $u$ is associated to EBS $b$ using channel $c$ at slot $t$. Constraints~\eqref{cns:6} are non-linear. These kind of constraints are known as indicator constraints (ICs), which use binary variables to control whether some linear constraints are active or not. In~\eqref{cns:6}, the binary variable $x_{u,b,t}^c$ is used to control whether user $u$	satisfies its required slots $\nu_{u,b}^c$ or not. ICs can be easily modeled in modern solvers such as CPLEX~\cite{cplex}. Nonetheless, ICs can be theoretically linearized using the bigM method~\cite{Schrijver:1986:TLI:17634}. Therefore, Constraints~\eqref{cns:5} can be rewritten as
	\begin{align}
	&\sum_{s\in\X{T}}x_{u,b,s}^c\geqslant \nu_{u,b}^cx_{u,b,t}^c,\quad\forall(u,b,c,t)\in\X{U}\times\X{B}\times\X{C}\times\X{T}.\tag{P1i}\label{cns:6:1}\\
	&\sum_{s\in\X{T}}x_{u,b,s}^c\leqslant \nu_{u,b}^cx_{u,b,t}^c+M(1-x_{u,b,t}^c),\quad\forall(u,b,c,t)\in\X{U}\times\X{B}\times\X{C}\times\X{T},\label{cns:6:2}\tag{P1j}
	\end{align}
	where $M$ is a large positive number.
	
	We can see from~\eqref{cns:6:1}, that if $x_{u,b,t}^c=1$, then $\sum_{s\in\X{T}}x_{u,b,s}^c=\nu_{u,b}^c,\forall(u,b,c,t)\in\X{U}\times\X{B}\times\X{C}\times\X{T}.$
	Also, if $x_{u,b,t}^c=0$, then $0\leqslant\sum_{s\in\X{T}}x_{u,b,s}^c\leqslant M,\forall(u,b,c,t)\in\X{U}\times\X{B}\times\X{C}\times\X{T},$ which is obviously
	true since $M$ is chosen large enough. It is clear that choosing $M=T$ is sufficient.
	
	\item The constraints that guarantee the deadline of the users can be expressed as
	\begin{align}\label{cns:7}
	d_u\geqslant t x_{u,b,t}^c,\quad\forall(u,b,c,t)\in\X{U}\times\X{B}\times\X{C}\times\X{T}.\tag{P1k}
	\end{align}
\end{itemize}

The objective of \textit{RAED} is to maximize the number of served users. Hence, the objective function can be  written as
\begin{align}\label{obj}
\sum_{u\in\X{U}}\sum_{b\in\X{B}}\sum_{c\in\X{C}}\sum_{t\in\X{T}}x_{u,b,t}^c/\nu_{u,b}^c.
\end{align}

With that said, we can formulate \textit{RAED} as the following ILP.
\begin{problem}\label{pb:2}
	\begin{alignat}{2}
	& \maximize_{[x_{u,b,t}^c],[z_{b,t}]} &\qquad &\sum_{u\in\X{U}}\sum_{b\in\X{B}}\sum_{c\in\X{C}}\sum_{t\in\X{T}}x_{u,b,t}^c/\nu_{u,b}^c\label{obj:ASE:2}\\
	& \subjto
	& & x_{u,b,t}^c\in\{0,1\},z_{b,t}\geqslant0,\quad\forall(u,b,c,t)\in\X{U}\times\X{B}\times\X{C}\times\X{T},\label{cns:1:ASE:2}\\
	& & & \eqref{cns:1}-\eqref{cns:7}\notag.
	\end{alignat}
\end{problem}

Using ILP-based solvers, we can optimally solve (not necessarily in polynomial-time) \textit{RAED} by solving~\eqref{pb:2}. In the next subsection, we analyze the computational complexity of \textit{RAED} in different cases. We refer to \textit{RAED} in case $X$ as \textit{RAED-X}. The proposed algorithm to solve \textit{RAED-X} is denoted as \texttt{ALG-X} and the optimal algorithm (obtained by solving~\ref{pb:2}) is denoted as \texttt{OPT-X}.

\subsection{Complexity Analysis}\label{subsec:NPhard}
In order to characterize the computational complexity of \textit{RAED}, we consider four cases: (i) single channel and single EBS (SCSB), (ii) single channel and multiple EBSs (SCMB), (iii) multiple channels and single EBS (MCSB), and (iv) multiple channels, multiple EBSs (MCMB). We summarize the results of our analysis in table~\ref{table:cc} where $\mathscr{P}$ and $\mathscr{NP}$ denotes the polytime and the nondeterministic polytime complexity classes, respectively. Table~\ref{table:cc} presents also the proposed algorithms along with their worst-case running-time complexities. We define $L\coloneq\min\{U,B\}$. 
\begin{table}[htb!]
	\centering
	\setlength\tabcolsep{0pt}
	\caption{Complexity Classification}
	\label{table:cc}
	\adjustbox{max width=\columnwidth}{
		\begin{tabular*}{\linewidth}{@{\extracolsep{\fill}} p{0.07\textwidth} p{0.22\textwidth} p{0.22\textwidth} p{0.22\textwidth} p{0.22\textwidth} }
			\toprule
			\multirow{2.2}{*}{}
			& \multicolumn{4}{c}{Cases}\\
			\cmidrule{2-5}
			& \textit{RAED-SCSB} & \textit{RAED-SCMB} & \textit{RAED-MCSB} & \textit{RAED-MCMB}\\
			\midrule
			Class & $\mathscr{P}$ & $\mathscr{NP}$-hard & $\mathscr{NP}$-hard & $\mathscr{NP}$-hard\\ \midrule
			Alg. & \begin{enumerate}[nosep, leftmargin=*,before=\vspace{-0.6\baselineskip},after=\vspace{-\baselineskip}] 
				\item \texttt{ALG-SCSB$_1$}
				\item \texttt{ALG-SCSB$_2$}
			\end{enumerate}
			& \texttt{ALG-SCMB} & \texttt{ALG-MCSB} & \texttt{ALG-MCMB}  \\  \midrule
			Type & \begin{enumerate}[nosep, leftmargin=*,before=\vspace{-0.6\baselineskip},after=\vspace{-\baselineskip}] 
				\item Optimal
				\item Optimal
			\end{enumerate}
			& Approximation & Heuristic & Heuristic  \\  \midrule 
			Time & \begin{enumerate}[nosep, leftmargin=*,before=\vspace{-0.6\baselineskip},after=\vspace{-\baselineskip}] 
				\item $O(UT^2+TU^2)$
				\item $O(T^2+U\log U)$
			\end{enumerate} & $O(BL(UT^2+TU^2))$ & $O(U^2T+UCT^2)$ & $O(BL(U^2T+UCT^2))$ \\
			\bottomrule
	\end{tabular*}}
\end{table}

The first case of \textit{RAED-SCSB} is considered in section~\ref{sec:SAED:BS} where it is shown that it belongs to the class $\X{P}$, i.e. the polynomial-time (polytime) complexity class. 

The second case of \textit{RAED-SCMB} is analyzed in the following. 
\begin{lemma}\label{lemma:1}
	\textit{RAED-SCMB} is $\X{NP}$-hard.
\end{lemma}
\begin{IEEEproof}
	We show that a special case of \textit{RAED-SCMB} is $\X{NP}$-hard. Precisely, when all deadlines are identical and energy is always available, \textit{RAED-SCMB} is still $\X{NP}$-hard. An instance of \textit{GAP} is given by a set of items and a set of bins, where a weight $w_{u,b}$ and a profit $p_{u,b}$ are given for each item $u$ and bin $b$. Further each bin $b$ has capacity $W_b$. The objective of \textit{GAP} is to maximize the profit of the items packed into the bins while respecting the capacity of the used bins. It is known that the special case of \textit{GAP} where $p_{u,b}=1$ is also $\X{NP}$-hard~\cite{Chekuri:713:728}. \textit{GAP} is reduced, in polytime, to \textit{RAED-SCMB} as follows. Given an instance of \textit{GAP} with $p_{u,b}=1$, the set of users corresponds to the set of items, the set of EBSs corresponds to the set of bins, $\nu_{u,b}$ corresponds to $w_{u,b}$, and the budget of each EBS $b$ is $W_b=T$. It is easy to see that this instance of \textit{RAED-SCMB} is solved if and only if \textit{GAP} is solved. Therefore, an algorithm that solves \textit{RAED-SCMB} in polytime can be used to solve GAP in polytime, which is not possible unless $\X{P}=\X{NP}$. This proves that \textit{RAED-SCMB} must be also $\X{NP}$-hard and thereby proves the lemma.
\end{IEEEproof}

The case of \textit{RAED-MCSB} is analyzed next.  
\begin{lemma}\label{lm:np2}
	\textit{RAED-MCSB} is $\X{NP}$-hard. 
\end{lemma}
\begin{IEEEproof}
	We consider \textit{RAED-MCSB} when $C=2$, the deadlines are common and there is enough energy available at the EBS. Thus, \textit{RAED-MCSB} is defined with $B=1$, $A_{1,t}=T$, $d_u=T$ for all $u$, and $C=2$. In this case, we are given one EBS that has enough energy across all slots, a set of users $\{1,2,\ldots,U\}$ and two channels, where each user $u$ requires $\nu_{u}^c$ slots using channel $c=1$ or $c=2$. We are trying to maximize the number of scheduled users at the EBS while respecting the common deadline $T$. We reduce in polytime \textit{PARTITION}~\cite{Garey:1979} to a decision version of \textit{RAED-MCSB}. In \textit{PARTITION} we are given a set of positive integers $\X{S}=\{a_1,\ldots,a_S\}$ and we are asked to partition it into two disjoint sets $\X{S}_1$ and $\X{S}_2$ such that $\X{S}_1\cup\X{S}_2=\X{S}$ and $\sum_{i\in\X{S}_1}a_i=\sum_{i\in\X{S}_2}a_i$. Given an instance of \textit{PARTITION}, we construct an instance of a decision version of \textit{RAED-MCSB} as follows. Let $T=\lfloor\sum_{i\in\X{S}}a_i/2\rfloor$ (assume without loss of generality that $T=\sum_{i\in\X{S}}a_i/2$) and $\nu_u^1=\nu_u^2=a_u$ for all $u$. The decision version of \textit{RAED-MCSB} is: Given this instance, can we schedule all users at the EBS?
	
	On the one hand, if we can solve \textit{PARTITION}, then all elements of $\X{S}_1$ (resp. $\X{S}_2$) can be scheduled at the beginning of the frame using channel $1$ (resp. channel $2$). Thus, the decision version of \textit{RAED-MCSB} is solved. On the other hand, if we can schedule all users using the two channels, then the users scheduled using channel $1$ (resp. channel $2$) can be chosen to represent the elements of $\X{S}_1$ (resp. $\X{S}_2$). It is clear that $\X{S}_1\cup\X{S}_2=\X{S}$. Also, since we scheduled all users and $T=\sum_{i\in\X{S}}a_i/2$, then the users scheduled using channel $1$ (or channel $2$) cannot require more nor less than $\sum_{i\in\X{S}}a_i/2$. Therefore, the users scheduled using channel $1$ (or channel $2$) require exactly $\sum_{i\in\X{S}}a_i/2$. Consequently, $\sum_{i\in\X{S}_1}a_i=\sum_{i\in\X{S}_2}a_i=\sum_{i\in\X{S}}a_i/2$. Thus, \textit{PARTITION} is solved. 
	
	We can see that the created instance of the decision version of \textit{RAED-MCSB} is done in polytime and hence it is $\X{NP}$-hard. This proves the lemma.
\end{IEEEproof}

\begin{remark}
	\textit{RAED-MCSB} can be solved in polytime in a very restricted case. Specifically, when $C\geqslant U$ and each user uses a different channel from every other user and requires the minimum number of slots from the allocated channel, thus, all users can be scheduled at the same slots. Therefore, the following channel allocation is optimal in terms of maximizing the number of scheduled users: for each user $u$, find the channel $c_u$ such that $\nu_{u}^{c_u}=\min_{c\in\X{C}}\nu_{u}^c$. Once the channel allocation is obtained, the problem is reduced to \textit{RAED-SCSB}, which can be solved by applying the proposed algorithm discussed in section~\ref{sec:SAED:BS}.
\end{remark}

Finally, the fourth case of \textit{RAED-MCMB} can be shown to be $\X{NP}$-hard based on the previous results of lemma~\ref{lemma:1} and lemma~\ref{lm:np2}.


In order to solve \textit{RAED} efficiently, we propose polytime, approximation, or heuristic algorithms depending on the case. We start by studying \textit{RAED} with single channel (i.e., $C=1$), which represents \textit{RAED-SCSB} and \textit{RAED-SCMB}. In these two cases, \textit{RAED} involves only user scheduling and association. 

\section{\textit{RAED} in the Case of Single Channel}\label{sec:SAED:BS}
\subsection{\textit{RAED-SCSB}}
\subsubsection{Arbitrary Deadlines} This subsection considers \textit{RAED-SCSB} when the deadlines of the users are arbitrary. Starting by solving \textit{RAED-SCSB} is important as it helps characterizing the structure of the solution in the more general cases. 


First, we introduce the following definition.
\begin{definition}[A schedule]\ \\
  \indent A schedule $\Sigma_{\text{EBS}}=[\sigma_1,\sigma_2,\ldots,\sigma_T]$ of the EBS is an allocation of a set of users $\{1,2,\ldots,U\}$ to a set of slots $\{1,2,\ldots,T\}$. Here, if $\sigma_t=u$, then user $u$ is scheduled at slot $t$ (we say that slot $t$ is busy), and if $\sigma_t=0$, then slot $t$ is idle (not busy). A preemptive schedule is one where the transmission of some users are interrupted and resumed later on. A non-preemptive schedule is one that is not preemptive. See Fig.~\ref{example:schedule} for an example.
\end{definition}
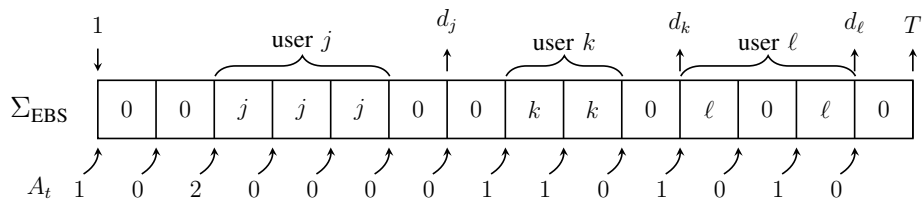
\begin{figure}[ht!]
	\centering
	\resizebox{.7\textwidth}{!}{%
		\begin{tikzpicture}[thick]
		\draw (0,0) node{} rectangle (14,1) {};
		\draw (1,0) -- (1,1);
		\draw (2,0) -- (2,1);
		\draw (3,0) -- (3,1);
		\draw (4,0) -- (4,1);
		\draw (5,0) -- (5,1);
		\draw (6,0) -- (6,1);
		\draw (7,0) -- (7,1);
		\draw (8,0) -- (8,1);
		\draw (9,0) -- (9,1);
		\draw (10,0) -- (10,1);
		\draw (11,0) -- (11,1);
		\draw (12,0) -- (12,1);
		\draw (13,0) -- (13,1);
		\draw (14,0) -- (14,1);
		\node (E01) at (0,0) {};
		\node (E02) at (-.4,-.7) {};
		\node (E11) at (1,0) {};
		\node (E12) at (0.6,-.7) {};
		\node (E21) at (2,0) {};
		\node (E22) at (1.6,-.7) {};
		\node (E31) at (3,0) {};
		\node (E32) at (2.6,-.7) {};
		\node (E41) at (4,0) {};
		\node (E42) at (3.6,-.7) {};
		\node (E51) at (5,0) {};
		\node (E52) at (4.6,-.7) {};
		\node (E61) at (6,0) {};
		\node (E62) at (5.6,-.7) {};
		\node (E71) at (7,0) {};
		\node (E72) at (6.6,-.7) {};
		\node (E81) at (8,0) {};
		\node (E82) at (7.6,-.7) {};
		\node (E91) at (9,0) {};
		\node (E92) at (8.6,-.7) {};
		\node (E101) at (10,0) {};
		\node (E102) at (9.6,-.7) {};
		\node (E111) at (11,0) {};
		\node (E112) at (10.6,-.7) {};
		\node (E121) at (12,0) {};
		\node (E122) at (11.6,-.7) {};
		\node (E131) at (13,0) {};
		\node (E132) at (12.6,-.7) {};
		\node (E141) at (14,0) {};
		\node (E142) at (13.6,-.7) {};
		\node (C1) at (0,1) {};
		\node (C1) at (0,1) {};
		\node (D1) at (0,1.7) {};
		\node (C2) at (14,1) {};
		\node (D2) at (14,1.7) {};
		\node (C3) at (13,1) {};
		\node (D3) at (13,1.7) {};
		\node (C4) at (10,1) {};
		\node (D4) at (10,1.7) {};
		\node (C6) at (6,1) {};
		\node (D6) at (6,1.7) {};
		\node at (.5,.5) {$0$};
		\node at (1.5,.5) {$0$};
		\node at (2.5,.5) {$j$};
		\node at (3.5,.5) {$j$};
		\node at (4.5,.5) {$j$};
		\node at (5.5,.5) {$0$};
		\node at (6.5,.5) {$0$};
		\node at (7.5,.5) {$k$};
		\node at (8.5,.5) {$k$};
		\node at (9.5,.5) {$0$};
		\node at (10.5,.5) {$\ell$};
		\node at (11.5,.5) {$0$};
		\node at (12.5,.5) {$\ell$};
		\node at (13.5,.5) {$0$};
		\draw [thick,decorate,decoration={brace,amplitude=10pt},xshift=0.4pt,yshift=0.6pt](2,1.05) -- (5,1.05)
		node[black,midway,yshift=0.6cm] {user $j$};
		\draw [thick,decorate,decoration={brace,amplitude=10pt},xshift=0.4pt,yshift=0.6pt](7,1.05) -- (9,1.05)
		node[black,midway,yshift=0.6cm] {user $k$};
		\draw [thick,decorate,decoration={brace,amplitude=10pt},xshift=0.4pt,yshift=0.6pt](10,1.05) -- (13,1.05)
		node[black,midway,yshift=0.6cm] {user $\ell$};
		\path[->,>=stealth] (D1) edge node [above,label=above:{$1$}] {} (C1);
		\path[->,>=stealth] (C2) edge node [above,label=above:{$T$}] {} (D2);
		\path[->,>=stealth] (C3) edge node [above,label=above:{$d_\ell$}] {} (D3);
		\path[->,>=stealth] (C4) edge node [above,label=above:{$d_k$}] {} (D4);
		\path[->,>=stealth] (C6) edge node [above,label=above:{$d_j$}] {} (D6);
		\path[->,>=stealth] (E02) edge[bend right] node [right,label=below left:{$1$}] {} (E01);
		\path[->,>=stealth] (E12) edge[bend right] node [right,label=below left:{$0$}] {} (E11);
		\path[->,>=stealth] (E22) edge[bend right] node [right,label=below left:{$2$}] {} (E21);
		\path[->,>=stealth] (E32) edge[bend right] node [right,label=below left:{$0$}] {} (E31);
		\path[->,>=stealth] (E42) edge[bend right] node [right,label=below left:{$0$}] {} (E41);
		\path[->,>=stealth] (E52) edge[bend right] node [right,label=below left:{$0$}] {} (E51);
		\path[->,>=stealth] (E62) edge[bend right] node [right,label=below left:{$0$}] {} (E61);
		\path[->,>=stealth] (E72) edge[bend right] node [right,label=below left:{$1$}] {} (E71);
		\path[->,>=stealth] (E82) edge[bend right] node [right,label=below left:{$1$}] {} (E81);
		\path[->,>=stealth] (E92) edge[bend right] node [right,label=below left:{$0$}] {} (E91);
		\path[->,>=stealth] (E102) edge[bend right] node [right,label=below left:{$1$}] {} (E101);
		\path[->,>=stealth] (E112) edge[bend right] node [right,label=below left:{$0$}] {} (E111);
		\path[->,>=stealth] (E122) edge[bend right] node [right,label=below left:{$1$}] {} (E121);
		\path[->,>=stealth] (E132) edge[bend right] node [right,label=below left:{$0$}] {} (E131);
		\node at (-1,-.85) {$A_t$};
		\node at (-1,.5) {\large $\Sigma_{\text{EBS}}$};
		\end{tikzpicture}
	}
	\caption{\footnotesize An example of one EBS and its corresponding preemptive schedule $\Sigma_{\text{EBS}}$.}
	\label{example:schedule}
\end{figure}

It is to be noticed that when energy is always available, \textit{RAED-SCSB} becomes equivalent to maximizing the number of early jobs in a single machine~\cite{Pinedo:2016}, which can be solved optimally using Moore-Hodgson's algorithm~\cite{Pinedo:2016} that uses a carefully-modified version of the earliest deadline first (\texttt{EDF}) scheduling rule.

To solve \textit{RAED-SCSB}, we propose a polytime optimal algorithm, called \texttt{ALG-SCSB$_1$}, which schedules the maximum number of users while respecting the energy and deadlines constraints. Energy constraints state that users are scheduled only when energy is available and that they should be scheduled for their required number of slots. Deadline constraints state that the scheduled users cannot miss their deadlines. First, we describe \texttt{ALG-SCSB$_1$}, then we prove its optimality.

\paragraph{Description of \texttt{ALG-SCSB$_1$}}
Before going into the details, we start by the following notations and definitions. Lowercase and boldface letters denote vectors whereas uppercase and boldface letters denote matrices. A set and its cardinality are denoted by the same calligraphic and italic letter, respectively. For example, $\X{S}$ denotes a set and $S$ denotes its cardinality. All sets are ordered sets, i.e., the $i$th element of $\{a_1,a_2,\ldots,a_S\}$ is $a_i$. A matrix $\mathbf{A}$ is sometimes denoted by $[a_{ij}]$.

\begin{definition}[A feasible schedule]~\newline
  \indent A schedule is called \textit{feasible} if it is energy-feasible and deadline-feasible. It is energy-feasible if the scheduled users meet the energy constraints and it is deadline-feasible if the
  scheduled users meet the deadlines constraints.
\end{definition}

\begin{definition}[An $\ell$-optimal schedule]~\newline
  \indent A schedule $\Sigma^\ell=[\sigma_1^\ell,\sigma_2^\ell,\ldots,\sigma_T^\ell]$ is called $\ell$-optimal, if it is a feasible schedule of the users from $\{1, 2, \ldots, \ell\}$ and it schedules the maximum number of users from $\{1, 2, \ldots, \ell\}$.
\end{definition}

\begin{algorithm}[tbp]
  \caption{Optimal algorithm for \textit{RAED-SCSB}}
  \label{alg:mhe}
  \begin{algorithmic}[1]
    \Function{\texttt{ALG-SCSB$_1$}}{$\X{U},\mathbf{d},\mathbf{A},\pmb{\nu}$}
    	\State Sort the users according to \texttt{EDF}
    	\State Set $\Sigma$ to an empty (idle) schedule, $\Sigma\gets[0,\ldots,0]$; $u\gets1$ and $\X{S}^0\gets\emptyset$
    	\For{$u\gets1$ \textbf{to} $U$}\label{mhe:line7}
    		\State Set $\X{S}^{u}\gets\X{S}^{u-1}\cup\{u\}$\label{mhe:line8}
    		\State Set $x\gets0$, $t\gets1$, and $r\gets1$\label{mhe:line10}
    		\While{$t\leqslant T$}\label{mhe:line12}
    			\State Set $r\gets\max\{r,t\}$\label{mhe:line121}
    			\If{$A_t>0$}
    				\State Set $\delta\gets\min\{A_t,\nu_u-x,T-r+1\}$
    				\For{$s$ \textbf{in} $\{r,\ldots,r+\delta-1\}$}
    					\If{$\sigma_s=0$}
    						\State Schedule user $u$ at $s$, $\sigma_s\gets u$\label{mhe:line13}
				    		\State Set $x\gets x+1$ and $r\gets r+1$
				    		\State Set $A_t\gets A_t-1$
			    		\EndIf
			    	\EndFor
			    \Else
				    \State Set $t\gets t+1$\label{mhe:line37}
    			\EndIf
			    \If{$\nu_u=x$ \textbf{and} $d_u\geqslant r-1$}\label{mhe:line23}
				    \State \textbf{break}\label{mhe:line24}
			    \ElsIf{$\nu_u=x$ \textbf{and} $d_u<r-1$}\label{mhe:line25}
				    \State Find the largest user $\ell$, $\ell\gets\argmax\{\nu_i:i\in\X{S}^u\}$\label{mhe:line26}
				    \State $(\Sigma,\mathbf{A},\X{S}^u,t,c)\gets$ \texttt{UPDATE}($\Sigma,\mathbf{A},\X{S}^u,\ell$)\label{mhe:line27}
				    \State \textbf{break}\label{mhe:line28}
			    \ElsIf{$\nu_u>x$ \textbf{and} $\max\{r,t\}>T$}\label{mhe:line29}
				    \State Find the largest user $\ell$, $\ell\gets\argmax\{\nu_i:i\in\X{S}^u\}$\label{mhe:line30}
				    \State $(\Sigma,\mathbf{A},\X{S}^u,t,c)\gets$ \texttt{UPDATE}($\Sigma,\mathbf{A},\X{S}^u,\ell$)\label{mhe:line31}
				    \If{$\ell=u$}\label{mhe:line32}
					    \State \textbf{break}\label{mhe:line33}
				    \EndIf\label{mhe:line34}
			    \EndIf
		    \EndWhile\label{mhe:line39}
	    \EndFor
    	\State\Return ($\Sigma$, $\X{S}^U$)
	    \EndFunction
  \end{algorithmic}
\end{algorithm}

\texttt{ALG-SCSB$_1$} is described in the pseudo-code shown in Algorithm~\ref{alg:mhe}. It works as follows. First, it  sorts the users according to \texttt{EDF}. Then, it starts with an empty (idle) schedule $\Sigma$ (i.e., $\Sigma=[0,0,\ldots,0]$) and it iterates the set of sorted users while checking the energy and deadlines constraints. Let $\X{S}^{u-1}$ be the set of users already scheduled in $\Sigma$ before the start of the $u$th iteration. At the $u$th iteration, user $u$ is the one being scheduled and hence \texttt{ALG-SCSB$_1$} adds it to $\X{S}^{u-1}$, i.e., \texttt{ALG-SCSB$_1$} creates the set $\X{S}^u=\X{S}^{u-1}\cup\{u\}$. Next, \texttt{ALG-SCSB$_1$} iterates the slots $t=1,2,\ldots,T$ and schedules $u$ at some slot $s$ whenever $A_t>0$ for $t\leqslant s$. User $u$ is scheduled at slot $s$ only if $s$ is an idle slot ($\sigma_{s}=0$). If $A_t=0$, then the next slot is considered. Note that after scheduling user $u$, \texttt{ALG-SCSB$_1$} goes through three \textbf{if} conditions in which it checks the energy and the deadlines constraints. These conditions are given in the following list. 
\begin{enumerate}
	\item User $u$ is allocated $\nu_u$ slots and its deadline $d_u$ is respected. In this case, the current user is skipped and the algorithm goes to user $u+1$.
	\item User $u$ is allocated $\nu_u$ slots but its deadline $d_u$ is not respected. In this case, \texttt{ALG-SCSB$_1$} removes from $\X{S}^u$ the largest user $\ell$, updates $\Sigma$ and $\mathbf{A}$, and goes to user $u+1$.
	\item User $u$ is not allocated $\nu_u$ slots yet (because there is no enough energy) and there is no time left in the frame. In this case, \texttt{ALG-SCSB$_1$} removes from $\X{S}^u$ the largest user $\ell$, updates $\Sigma$ and $\mathbf{A}$, and goes to user $u+1$ only if $\ell=u$. In other words, \texttt{ALG-SCSB$_1$} goes to user $u+1$ if it removed user $u$, but continues on scheduling $u$ otherwise.
\end{enumerate}
In the last two conditions, a rescheduling procedure called \texttt{UPDATE} is invoked. It mainly performs three operations: (1) removes $\ell$ from $\X{S}^u$, (2) shifts all users scheduled after $\ell$ to the left, and (3) updates the slots (energy) $\mathbf{A}$. Algorithm~\ref{alg:reschedule} summarizes the rescheduling procedure. The last two operations of the rescheduling procedure are discussed next. On the one hand, the second operation mainly finds the sets of slots $\X{T}_1$ and $\X{T}_2$, where $\X{T}_1$ represents the slots during which $\ell$ is scheduled whereas $\X{T}_2$ is the set of slots during which all users $j\neq\ell$ are scheduled after $\ell$, i.e. $\X{T}_1=\{t\in\X{T}:\sigma_{t}=\ell\}$ and $\X{T}_2=\{t\in\X{T}:\sigma_{t}\neq0\text{ and }\sigma_{t}\neq\ell\text{ and } t\geqslant\min\X{T}_1\}$. Next, for each slot $t\in\X{T}_2$, $\sigma_t$ is shifted to the left either to a slot of $\X{T}_1$ or to a slot of $\X{T}_2$ of an already shifted user. Of course, each time a user $u$ is shifted, we must guarantee that it is scheduled at some idle slot where energy is available (its deadline will be respected since it will be shifted to the left). Removing user $\ell$ and shifting all subsequent users update the schedule $\Sigma$ by creating new idle slots in it. On the other hand, the third operation calculates the set of new idle slots $\X{T}_3$ in $\Sigma$. For each slot $t\in\X{T}_3$, some user in $\Sigma$ was scheduled at $t$ and hence it was allocated one unit of energy at that slot from an earlier slot $s\leqslant t$. Thus, for each slot $t\in\X{T}_3$, the third operation calculates the time slot $s$ from which one unit of energy (from $A_{s}$) was used at slot $t$ and it updates $A_s$ as $A_s\gets A_s+1$. 

After performing these operations, \texttt{ALG-SCSB$_1$} returns the tuple $(\Sigma,\mathbf{A},\X{S}^u,t,r)$. In this tuple, $t$ represents the current time iteration of the \textbf{while} loop of \texttt{ALG-SCSB$_1$}, $r$ is the time where the next user will be scheduled, $\Sigma$ is the new schedule and $\mathbf{A}$ is the updated energy.
\begin{algorithm}[ht!]
	\caption{The rescheduling procedure}
	\label{alg:reschedule}
	\begin{algorithmic}[1]
		\Function{\texttt{UPDATE}}{$\Sigma,\mathbf{A},\X{S}^u,\ell$}
		\State $\X{S}^u\gets\X{S}^u\backslash\{\ell\}$
		\State Update $\Sigma$ by shifting to the left the users scheduled after $\ell$\label{upd:3}
		\State Update the allocated slots $\mathbf{A}$
		\State $t\gets1$ and $r\gets1$
		\State\Return $(\Sigma,\mathbf{A},\X{S}^u,t,r)$
		\EndFunction
	\end{algorithmic}
\end{algorithm}

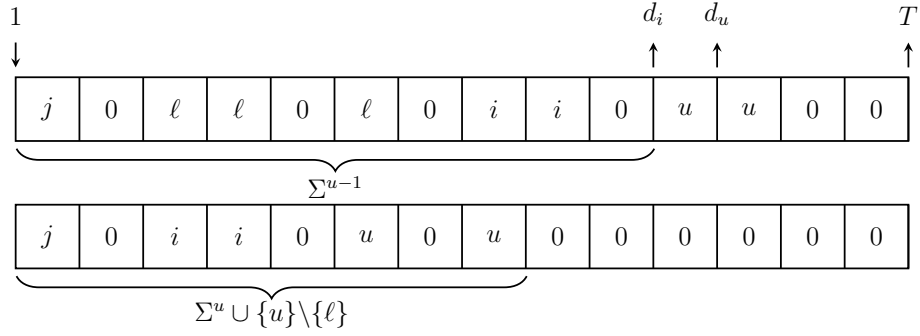
\begin{figure}[ht!]
  \centering
  \resizebox{.7\textwidth}{!}{%
    \begin{tikzpicture}[thick]
      \draw (0,0) node{} rectangle (14,1) {};
      \draw (1,0) -- (1,1);
      \draw (2,0) -- (2,1);
      \draw (3,0) -- (3,1);
      \draw (4,0) -- (4,1);
      \draw (5,0) -- (5,1);
      \draw (6,0) -- (6,1);
      \draw (7,0) -- (7,1);
      \draw (8,0) -- (8,1);
      \draw (9,0) -- (9,1);
      \draw (10,0) -- (10,1);
      \draw (11,0) -- (11,1);
      \draw (12,0) -- (12,1);
      \draw (13,0) -- (13,1);
      \draw (14,0) -- (14,1);
      \node (C1) at (0,1) {};
      \node (D1) at (0,1.7) {};
      \node (C2) at (14,1) {};
      \node (D2) at (14,1.7) {};
      \node (C3) at (11,1) {};
      \node (D3) at (11,1.7) {};
      \node (C4) at (10,1) {};
      \node (D4) at (10,1.7) {};
      \node at (.5,.5) {$j$};
      \node at (1.5,.5) {$0$};
      \node at (2.5,.5) {$\ell$};
      \node at (3.5,.5) {$\ell$};
      \node at (4.5,.5) {$0$};
      \node at (5.5,.5) {$\ell$};
      \node at (6.5,.5) {$0$};
      \node at (7.5,.5) {$i$};
      \node at (8.5,.5) {$i$};
      \node at (9.5,.5) {$0$};
      \node at (10.5,.5) {$u$};
      \node at (11.5,.5) {$u$};
      \node at (12.5,.5) {$0$};
      \node at (13.5,.5) {$0$};
      \draw [thick,decorate,decoration={brace,amplitude=10pt,mirror},xshift=0.4pt,yshift=-0.6pt](0,-.1) -- (10,-.1)
      node[black,midway,yshift=-0.6cm] {$\Sigma^{u-1}$};
      \path[->,>=stealth] (D1) edge node [above,label=above:{$1$}] {} (C1);
      \path[->,>=stealth] (C2) edge node [above,label=above:{$T$}] {} (D2);
      \path[->,>=stealth] (C3) edge node [above,label=above:{$d_u$}] {} (D3);
      \path[->,>=stealth] (C4) edge node [above,label=above:{$d_i$}] {} (D4);
      \draw (0,-2) node{} rectangle (14,-1) {};
      \draw (1,-2) -- (1,-1);
      \draw (2,-2) -- (2,-1);
      \draw (3,-2) -- (3,-1);
      \draw (4,-2) -- (4,-1);
      \draw (5,-2) -- (5,-1);
      \draw (6,-2) -- (6,-1);
      \draw (7,-2) -- (7,-1);
      \draw (8,-2) -- (8,-1);
      \draw (9,-2) -- (9,-1);
      \draw (10,-2) -- (10,-1);
      \draw (11,-2) -- (11,-1);
      \draw (12,-2) -- (12,-1);
      \draw (13,-2) -- (13,-1);
      \draw (14,-2) -- (14,-1);
      \node (C1) at (0,-1) {};
      \node (D1) at (0,-.3) {};
      \node (C2) at (14,-1) {};
      \node (D2) at (14,-.3) {};
      \node (C3) at (11,-1) {};
      \node (D3) at (11,-.3) {};
      \node (C4) at (10,-1) {};
      \node (D4) at (10,-.3) {};
      \node at (.5,-1.5) {$j$};
      \node at (1.5,-1.5) {$0$};
      \node at (2.5,-1.5) {$i$};
      \node at (3.5,-1.5) {$i$};
      \node at (4.5,-1.5) {$0$};
      \node at (5.5,-1.5) {$u$};
      \node at (6.5,-1.5) {$0$};
      \node at (7.5,-1.5) {$u$};
      \node at (8.5,-1.5) {$0$};
      \node at (9.5,-1.5) {$0$};
      \node at (10.5,-1.5) {$0$};
      \node at (11.5,-1.5) {$0$};
      \node at (12.5,-1.5) {$0$};
      \node at (13.5,-1.5) {$0$};
      \draw [thick,decorate,decoration={brace,amplitude=10pt,mirror},xshift=0.4pt,yshift=-0.6pt](0,-2.1) -- (8,-2.1)
      node[black,midway,yshift=-0.6cm] {$\Sigma^{u}\cup\{u\}\backslash\{\ell\}$};
    \end{tikzpicture}
  }
  \caption{\footnotesize An illustration of the proof of lemma~\ref{lemma:feasibility:mhe}.}
  \label{illustration:proof:lemma:feasibility:mhe}
\end{figure}

\paragraph{Optimality of \texttt{ALG-SCSB}}
Next, we prove that \texttt{ALG-SCSB$_1$} always returns an optimal schedule. First, we show that the returned schedule is feasible and then we prove its optimality. Let $\Sigma^u$ be the schedule found by \texttt{ALG-SCSB$_1$} at the end of iteration $u$, i.e., $\Sigma^u$ corresponds to the set $\X{S}^u$ of scheduled users. Throughout this part we apply set theory terminology to $\Sigma^{u}$, e.g., $\Sigma^{u-1}\cup\{u\}$ corresponds the set $\X{S}^{u-1}\cup\{u\}$ scheduled according to \texttt{ALG-SCSB$_1$}.
\begin{lemma}\label{lemma:feasibility:mhe}
  The schedule $\Sigma^u$ is feasible.
\end{lemma}
\begin{IEEEproof}\label{proof:lemma:feasibility:mhe}
  We prove the lemma by induction. For $u=1$, it is clear that $\Sigma^1$ is feasible, since $\Sigma^1$ either contains user $1$ or is empty and if it contains user $1$ then it is clear that it respects the enery and deadline constraints. Assume that $\Sigma^{u-1}$ is feasible. If $\Sigma^u=\Sigma^{u-1}\cup\{u\}$ (the rescheduling procedure is not invoked), then we are done. Otherwise (the rescheduling procedure is invoked), let $\ell$ be the user to be removed from $\Sigma^{u-1}\cup\{u\}$, i.e., $\ell$ is the user requiring the largest number of slots, called hereinafter \textit{the largest user}. We prove that
  $\Sigma^u=\Sigma^{u-1}\cup\{u\}\backslash\{\ell\}$ is feasible. We divide users in $\Sigma^u$ into three disjoint sets: (i) users scheduled before $\ell$, (ii) users scheduled after $\ell$, and (iii) user $u$. Notice that:
  \begin{itemize}
  \item[(i)] users scheduled before $\ell$ are left unchanged by the rescheduling procedure (user $j$ in Fig.~\ref{illustration:proof:lemma:feasibility:mhe});
  \item[(ii)] users scheduled after $\ell$ finish earlier due to the shifting operation in the rescheduling procedure. Hence, they respect their deadlines. Also, they respect the energy constraints since $\ell$ is the largest user and $\Sigma^{u-1}$ is energy-feasible (users $i$ and $k$ in Fig.~\ref{illustration:proof:lemma:feasibility:mhe}); and
  \item[(iii)] user $u$ respects the energy constraints because $\nu_u\leqslant\nu_\ell$ and $\Sigma^{u-1}$ is energy-feasible. Also, user $u$ respects its deadline because (1) the last user in $\Sigma^u$ finishes before the last user (say $i$) in $\Sigma^{u-1}$ due to the shifting operation and to the fact that $\ell$ is the largest user, (2) $d_u\geqslant d_i$ since users are sorted according to \texttt{EDF} and (3) $\Sigma^{u-1}$ is deadline-feasible. See Fig.~\ref{illustration:proof:lemma:feasibility:mhe}.
  \end{itemize}
  This proves that $\Sigma^u$ is deadline- and energy-feasible and therefore it is feasible. This proves the lemma.
\end{IEEEproof}

\begin{theorem}\label{theorem:optimality:mhe}
  The schedule $\Sigma^u$ is $u$-optimal for all $u=1,2,\ldots,U$.
\end{theorem}
\begin{IEEEproof}\label{proof:theorem:optimality:mhe}
  We first prove that for $\ell>u$, there exists an $\ell$-optimal schedule $\Sigma''$ that consists of users from $\Sigma^u$ and users from $\{u+1, u+2, \ldots, \ell\}$. We proceed by induction. Assume that there exists an $\ell$-optimal schedule $\Sigma'$ that consists of users from $\Sigma^{u-1}$ and users from $\{u, u+1, \ldots, \ell\}$. We have three cases:
  \begin{itemize}
  \item[(i)] If $\Sigma^u=\Sigma^{u-1}\cup\{u\}$. It is clear that an $\ell$-optimal schedule $\Sigma''$ exists (i.e., $\Sigma''=\Sigma'$) that consists of users from $\Sigma^u$ and users from $\{u+1, u+2, \ldots, \ell\}$.
  \item[(ii)] If $\Sigma^u=\Sigma^{u-1}\cup\{u\}\backslash\{j\}$ where $j\notin\Sigma'$. It is also clear that an $\ell$-optimal schedule $\Sigma''$ exists (i.e., $\Sigma''=\Sigma'$) that consists of users from $\Sigma^u$ and users from $\{u+1, u+2, \ldots, \ell\}$.
  \item[(iii)] If $\Sigma^u=\Sigma^{u-1}\cup\{u\}\backslash\{j\}$ where $j\in\Sigma'$. In this case, to obtain $\Sigma''$, we need to modify $\Sigma'$. We know that $\Sigma^{u-1}\cup\{u\}$ is not feasible. Hence, there exists $i\in\Sigma^{u-1}\cup\{u\}$ such that $i\notin\Sigma'$. Let $\Sigma''=\Sigma'\cup\{i\}\backslash\{j\}$. It is clear that $\Sigma''$ consists of users from $\Sigma^u$ and users from $\{u+1, u+2, \ldots, \ell\}$. Also, the number of users in $\Sigma'$ is the same as the number of users in $\Sigma''$. Hence, it remains to show that $\Sigma''$ is    feasible. We observe that $\Sigma''$ differs from $\Sigma'$ only on its intersection with users $\{1,2,\ldots,u\}$, since $i\in\Sigma^{u-1}\cup\{u\}$. We have $\Sigma''\cap\{1,2,\ldots,u\}\subseteq\Sigma^u$ and, by lemma~\ref{lemma:feasibility:mhe}, $\Sigma^u$ is feasible. Thus, $\Sigma''\cap\{1,2,\ldots,u\}$ is feasible. Based on the previous argument and the fact that \texttt{ALG-SCSB$_1$} removes always the largest user (i.e., $\nu_j\geqslant\nu_i$), we conclude that $\Sigma''$ is feasible. Finally, we have: for $\ell>u$, there exists an $\ell$-optimal schedule $\Sigma''$ that consists of users from schedule $\Sigma^u$ and users from $\{u+1,u+2, \ldots, \ell\}$.
  \end{itemize}
  We use the previous result and induction to prove that the schedule $\Sigma^u$ is $u$-optimal for all $u=1, 2, \ldots, U$. For $u=1$, it is clearly true that $\Sigma^1$ is $1$-optimal. Suppose that the $\Sigma^{u-1}$ is $u-1$-optimal. From the previous step (for $\ell=u>u-1$), we know that there exists a $u$-optimal schedule that consists of users from schedule $\Sigma^{u-1}$ and $\{u\}$. If $\Sigma^u=\Sigma^{u-1}\cup\{u\}$, then $\Sigma^u$ is $u$-optimal since, by assumption, $\Sigma^{u-1}$ is $u-1$-optimal. Otherwise, the $u$-optimal schedule (i) is a proper subset of users $\Sigma^{u-1}\cup\{u\}$ and (ii) has at least as many users as $\Sigma^{u-1}$ because
  the latter is $u-1$-optimal. Clearly, $\Sigma^u\cup\{u\}\backslash\{j\}$, where $j$ is the largest user, satisfies both conditions (i) and (ii) and hence it is $u$-optimal. Therefore, $\Sigma^u$ is $u$-optimal. This proves the theorem.
\end{IEEEproof}

We showed that \texttt{ALG-SCSB$_1$} returns an optimal solution to \textit{RAED-SCSB}. The worst-case complexity of
\texttt{ALG-SCSB$_1$} is $O(UT^2+U^2T+U\log U)=O(UT^2+U^2T)$, which is calculated as follows. First of all, it is clear that \texttt{ALG-SCSB$_1$} halts in a finite number of steps. The \textbf{for} loop of line~\ref{mhe:line7} halts in $O(U)$ steps. We show that the \textbf{while} loop of line~\ref{mhe:line12} halts in $O(T)$ steps. Let $t$ be an arbitrary iteration of this loop. We can see that if $A_t=0$, then the time is incremented. Hence, without loss of generality, assume that $A_t>0$ whenever \texttt{ALG-SCSB$_1$} enters this loop. With that said, we have three cases: (1) $\nu_u=x$ and $d_u\geqslant r-1$, (2) $\nu_u=x$ and $d_u<r-1$, or (3) $\nu_u>x$ and $\max\{r,t\}>T$. In the first two cases the \textbf{while} loop halts, since we break it in lines~\ref{mhe:line24} and~\ref{mhe:line28}, respectively. The last case is the most expensive one (in terms of time complexity). Whenever $\nu_u>x$ and $\max\{r,t\}>T$, \texttt{ALG-SCSB$_1$} calls the rescheduling procedure and goes to the next iteration starting from $t=1$. But, we can see that every time $A_t>0$, \texttt{ALG-SCSB$_1$} increases $x$ by at most $\delta$. Thus, at some iteration, $x$ will be greater than or equal to $\nu_u$. At this time, \texttt{ALG-SCSB$_1$} goes to one of the two previous \textbf{if} conditions and the \textbf{while} loop eventually halts. In the worst-case, $x$ is increased by $1$ every time, i.e., $\delta=1$. Hence, the \textbf{while} loop of line~\ref{mhe:line12} halts in $O(T)$ steps. Any iteration of the \textbf{while} loop has a worst-case complexity of $O(U+T)$---$O(U)$ to find the largest user and $O(T)$ to reschedule the users. Finally, the worst-case complexity  of \texttt{ALG-SCSB$_1$} is given by $O(UT(U+T)+U\log U)$. Therefore, \texttt{ALG-SCSB$_1$} halts in $O(UT^2+U^2T)$ steps in the worst-case.

\subsubsection{Common Deadlines}
This subsection solves \textit{RAED-SCSB} when users have common deadlines, i.e., $d_u=T$ for all $u\in\X{U}$. It proposes a less-complex algorithm than \texttt{ALG-SCSB$_1$}. This special case is called \textit{RAED-SCSB} with common deadlines (\textit{RAED-SCSB-COMMON}). When the energy is always available at the EBS, solving \textit{RAED-SCSB-COMMON} becomes straightforward. It can be reduced, in polytime, to an unweighted knapsack problem as follows. The users represent the items where the weight of item $u$ is given by the required slots $\nu_u$, the EBS represents the knapsack and the common deadline $T$ is the knapsack capacity. Hence, \textit{RAED-SCSB-COMMON} when the energy constraints are omitted can be solved optimally by sorting the users in increasing order of their required slots and scheduling the users one after the other until $T$ is reached. This procedure, named \texttt{PACK}, is applied at the beginning of the frame (i.e., at $t=1$) and accepts as inputs the set of users $\X{U}$, the common deadline $T$, the required number of slots $\pmb{\nu}$, and the time $t$. The worst-case time complexity of \texttt{PACK} is $O(T+U\log U)$, where $O(U\log U)$ steps are required to find the users to be scheduled and $O(T)$ steps are required to create the schedule. 

Now, to solve \textit{RAED-SCSB-COMMON} with energy constraints, we reduce it, in polytime, to an unweighted knapsack problem with cumulative capacity; that is, the capacity of the knapsack is not fixed to $T$ but changes from one slot to another depending on the energy arrivals. Therefore, knowing the maximum accumulated capacity, we can apply \texttt{PACK} to obtain an optimal solution to \textit{RAED-SCSB-COMMON}. Let $\Lambda_t$ be the accumulated capacity at time $t$, which is defined as the maximum number of slots that can be used by the EBS from slot $t$. The cumulative capacity procedure that calculates $\Lambda_t$ for $t\in\X{T}$ works iteratively as follows. In the first iteration, it calculates $\Gamma(t)\coloneq\sum_{i=1}^{t}A_{i}$. If no energy arrives during the period $\{t+1,\ldots,t+\Gamma(t)\}$, then no more than $\Gamma(t)$ slots can be used by the EBS and thus $\Lambda_t=\Gamma(t)$. Otherwise, $\Lambda_t>\Gamma(t)$. In fact, $\Lambda_t=\Gamma(t+\Gamma(t))$. Similarly, in the second iteration, if no energy arrives in the period $\{t'+1,\ldots,t+\Gamma(t')\}$, then $\Lambda_t=\Gamma(t')$. Otherwise, $\Lambda_t=\Gamma(t+\Gamma(t'))$ and the process continues this way. Of course the final accumulated capacity cannot exceed the remaining number of slots. Hence, $\Lambda_t=\min\left\{\Gamma(t'), T-t+1\right\}$, where $t'$ is the iteration such that no energy arrived in the period $\{t'+1,\ldots,t+\Gamma(t')\}$. This procedure is illustrated in Algorithm~\ref{algo:budget} and is called \texttt{BUDGET}.
\begin{algorithm}[ht!]
	\caption{The cumulative capacity procedure}
	\label{algo:budget}
	\begin{algorithmic}[1]
		\Function{\texttt{BUDGET}}{$\mathbf{A}$}
		\For{$t\gets1$ \textbf{to} $T$}
		\State $\Gamma(t)\gets\sum_{i=1}^{t}A_{i}$
		\State $t'\gets t+\Gamma(t)$
		\While{$t'\leqslant T$ \textbf{and} $t'\neq t+\Gamma(t')$}\label{budget:line1}
		\State $t'\gets t+\Gamma(t')$
		\EndWhile
		\State $\Lambda_{t}\gets\min\left\{\Gamma(t'), T-t+1\right\}$
		\EndFor
		\State $\Lambda^\star\gets\max\left\{\Lambda_1, \ldots, \Lambda_T\right\}$\label{budget:line10}
		\State $t^\star\gets\argmax\left\{\Lambda_1, \ldots, \Lambda_T\right\}$\label{budget:line11}
		\State\Return $(\Lambda^\star, t^\star)$
		\EndFunction
	\end{algorithmic}
\end{algorithm}
After calculating the accumulated capacity $\Lambda_t$ at each time $t$, \texttt{BUDGET} finds the maximum capacity $\Lambda^\star=\Lambda_{t^\star}=\max\left\{\Lambda_1, \ldots, \Lambda_T\right\}$ where $t^\star=\argmax\left\{\Lambda_1, \ldots, \Lambda_T\right\}$. Given $\Lambda^\star$ and $t^\star$, the algorithm that solves \textit{RAED-SCSB-COMMON} calls \texttt{PACK}$(\X{U}, \Lambda^\star,\pmb{\nu}; t=t^\star)$, which is applied at $t=t^\star$. Since $\Lambda^\star$ represents the maximum possible number of slots that can be used by the EBS, then this leads to an optimal solution to \textit{RAED-SCSB-COMMON}.

This algorithm, called \texttt{ALG-SCSB$_2$}, finds an optimal solution to \textit{RAED-SCSB-COMMON} in $O(T^2+T+U\log U)=O(T^2+U\log U)$ steps in the worst-case, where $O(T^2)$ steps are required to calculate the accumulated capacity $\Lambda^\star$ and $O(T+U\log U)$ steps are required to schedule the users by calling \texttt{PACK}. We can see that \texttt{ALG-SCSB$_2$} is much simple (less complex) than \texttt{ALG-SCSB$_1$} for arbitrary deadlines. This simplicity is obtained by exploiting the structure of the problem in which the users have common deadlines.

In the next section, we study \textit{RAED} in the single channel case but with multiple EBSs, i.e., \textit{RAED-SCMB}.
\subsection{\textit{RAED-SCMB}}\label{sec:SAED:MBS}
We showed in Section~\ref{sec:SAED:pbform} that \textit{RAED-SCMB} is $\X{NP}$-hard. Hence, we aim to design a polytime \textit{approximation algorithm} to solve \textit{RAED-SCMB}. Approximation algorithms~\cite{Williamson:2011:DAA:1971947} are algorithms that find ``good enough'' solutions very fast, where goodness is measured by the ratio of the value of the algorithm to the value of the optimal (e.g. brute-force) algorithm. In other words, approximation algorithms are polytime algorithms that are guaranteed theoretically to find solutions that are close to the optimal ones. 

\subsubsection{An Approximation Algorithm for \textit{RAED-SCMB}}\label{ssection:proposedalg}
To solve \textit{RAED-SCMB}, we propose \texttt{ALG-SCMB}---an iterative algorithm that uses \texttt{ALG-SCSB$_1$} as a subroutine. In short, for each iteration, \texttt{ALG-SCMB} finds the EBS that can schedule the maximum number of users. The pseudo-code of \texttt{ALG-SCMB} is given in Algorithm~\ref{alg}. We use the following additional notations. The collection of sets $\X{A}_{\X{B}}$ denotes $\{\X{A}_i: i\in\X{B}\}$, where $\X{A}_i$ is a set. Further, given a matrix $\mathbf{A}$, we use $\mathbf{A}_i$ to denote its $i$th row and $\mathbf{A}^i$ to denote its $i$th column.

\begin{algorithm}[ht!]
  \caption{Approximation algorithm for \textit{RAED-SCMB}}
  \label{alg}
  \begin{algorithmic}[1]
    \Function{\texttt{ALG-SCMB}}{$\X{U}, \X{B}, \pmb{\nu}, \mathbf{A}$}
    \State Set $\X{X}\gets\X{U}$, $\X{Y}\gets\X{B}$, $\X{Z}\gets\emptyset$, and $\X{A}\gets\emptyset$
    \While{$\X{X}\neq\emptyset$ \textbf{and} $\X{Y}\neq\emptyset$}\label{alg:line1}
    \For{$b$ \textbf{in} $\X{Y}$}\label{alg:4}
    \State $(\Sigma,\X{A}_{b})\gets$ \texttt{ALG-SCSB$_1$}$(\X{X},\mathbf{d},\mathbf{A}_{b},\pmb{\nu}^b)$
    \EndFor\label{alg:6}
    \State $b^\star\gets\argmax\left\{\X{A}_{\X{Y}}\right\}$
    \State $\X{Y}\gets\X{Y}\backslash\{b^\star\}$
    \State $\X{X}\gets\X{X}\backslash\X{A}_{b^\star}$
    \State $\X{Z}\gets\X{Z}\cup\{b^\star\}$
    \EndWhile
    \State\Return $\X{A}_{\X{Z}}$
    \EndFunction
  \end{algorithmic}
\end{algorithm}

In more details, \texttt{ALG-SCMB} works as follows. Let $\X{X}$ (resp. $\X{Y}$) be the set of remaining users (resp. the set of remaining EBSs) after any iteration. Initially, $\X{X}=\X{U}$ and $\X{Y}=\X{B}$. For each iteration, \texttt{ALG-SCMB} goes through $\X{Y}$ and schedules the users in $\X{X}$ according to \texttt{ALG-SCSB$_1$}. At the end of this iteration, \texttt{ALG-SCMB} creates a set $\X{A}_b$ of scheduled users to every EBS $b$. Next, it selects EBS $b^\star$ that schedules the maximum number of users $\X{A}_{b^\star}$. Then, it updates the set $\X{X}$ by removing the users $\X{A}_{b^\star}$ and updates the set $\X{Y}$ by removing EBS $b^\star$. At every iteration, \texttt{ALG-SCMB} updates the set $\X{Z}$ of used EBSs by adding EBS $b^\star$. And, when \texttt{ALG-SCMB} halts, it returns the tuple $\X{A}_{\X{Z}}$ of served users.

We can show that \texttt{ALG-SCMB} is a constant-factor approximation algorithm for \textit{RAED-SCMB}. This means that \texttt{ALG-SCMB} always returns a solution to \textit{RAED-SCMB} (i) in polytime, and (ii) whose value is bounded
below by a constant times the optimal value. The following lemma proves this result.

\begin{lemma}\label{theorem:1}
	\texttt{ALG-SCMB} is a $\frac{1}{2}$-approximation algorithm for \textit{RAED-SCMB}. 
\end{lemma}

\begin{IEEEproof}
We prove the lemma by showing that \texttt{ALG-SCMB} (i) halts in polytime, and (ii) the number of served users by \texttt{ALG-SCMB} is at least half the optimal number. The first point follows from the fact that \texttt{ALG-SCSB$_1$} is a polytime algorithm and that the \textbf{while} loop in line~\ref{alg:line1} of Algorithm~\ref{alg} halts in $L\coloneq\min(U,B)$ steps. Hence, \texttt{ALG-SCMB} halts in $O(BL(UT^2+U^2T))$ steps in the worst case. To prove the second point, let \texttt{OPT-SCMB} be an optimal algorithm for \textit{SAED-SCMB} and let $\X{O}_b$ be the set of users served by EBS $b$ in \texttt{OPT-SCMB} but which are not considered by any EBS in \texttt{ALG-SCMB}. Note that we do not have to know the elements of $\X{O}_b$ neither the optimal algorithm \texttt{OPT-SCMB}. According to our notations in Algorithm~\ref{alg}, $\X{A}_b$ corresponds to the set of users served by EBS $b$ in \texttt{ALG-SCMB}. Define $X_b\coloneq|\X{A}_b|$. Finally, let $\X{C}_b$ be the set of common users (i.e. users served by EBS $b$ in both \texttt{OPT-SCMB} and \texttt{ALG-SCMB}). The optimal number of served users in \texttt{OPT-SCMB} is $X^\star\coloneq\sum_{b=1}^{B}|\X{O}_b|+|\X{C}_b|$ whereas the number of served users in \texttt{ALG-SCMB} is $X\coloneq\sum_{b=1}^{B}X_b$. For any EBS $b$, all users in $\X{O}_b$ can be considered by \texttt{ALG-SCMB}. Since \texttt{ALG-SCMB} serves the maximum number of remaining users, then EBS $b$ must serve at least $|\X{O}_b|$, i.e., we have $X_b\geqslant|\X{O}_b|$, for all EBSs $b=1,2,\ldots,B$. Thus, summing over $b$, we obtain: 
\begin{align}\label{approx123}
	X\geqslant\sum_{b=1}^{B}|\X{O}_b|.
\end{align}

Now, there are two cases: whether or not the total number users served only by \texttt{OPT-SCMB} is greater than or equal to half the optimal number $X^\star$.
\begin{itemize}
	\item If $\sum_{b=1}^{B}|\X{O}_b|\geqslant X^\star/2$. In this case, based on~\eqref{approx123}, we have $X\geqslant X^\star/2$.
	\item If $\sum_{b=1}^{B}|\X{O}_b|\leqslant X^\star/2$. In this case, we have $\sum_{b=1}^{B}|\X{O}_b|+\sum_{b=1}^{B}|\X{C}_b|\leqslant X^\star/2+\sum_{b=1}^{B}|\X{C}_b|$ and hence $X^\star\leqslant X^\star/2 + \sum_{b=1}^{B}|\X{C}_b|$. Alternatively, $X^\star/2\leqslant \sum_{b=1}^{B}|\X{C}_b|$. Since the users in $\X{C}_b$ are served by both \texttt{ALG-SCMB} and \texttt{OPT-SCMB}, then $|\X{C}_b|\leqslant X_b$ for all $b$. Thus, $\sum_{b=1}^{B}|\X{C}_b|\leqslant X$. Finally, we obtain $X\geqslant X^\star/2$.
\end{itemize}
In both cases, we have $X^\star\geqslant X\geqslant X^\star/2$. Therefore, \texttt{ALG-SCMB} is a $\frac{1}{2}$-approximation algorithm for \textit{RAED-SCMB}. This proves the lemma.
\end{IEEEproof}

\begin{remark}
  The approximation factor of $1/2$ of \texttt{ALG-SCMB} is tight. That is, no better factor can be obtained for \texttt{ALG-SCMB}. This is shown in the following example. Let $U=4$, $B=2$, $T=2$, and $A_{11}=2$, $A_{12}=0$, $A_{21}=2$, $A_{22}=0$. The required slots are given by $\nu_{11}=\nu_{21}=\nu_{22}=\nu_{31}=\nu_{32}=\nu_{41}=1$ and $\nu_{42}=\nu_{12}=2$. On the one hand, the optimal algorithm associates users 1 and 4 to EBS 1 and schedules them at slots 1 and 2, respectively, and associates users 2 and 3 to EBS 2 and schedules them at slots 1 and 2, respectively. Hence, the optimal algorithm serves 4 users. On the other hand, if \texttt{ALG-SCMB} associates users 2 and 3 to EBS 1 and schedules them at slots 1 and 2, respectively, then it can only associate user 1 or user 4 to EBS 2 and schedules it at either slot 1 or 2. Hence, \texttt{ALG-SCMB} serves 2 users. We can see that $X=X^\star/2$. Therefore, the approximation factor $1/2$ of \texttt{ALG-SCMB} is tight.
\end{remark}

\section{\textit{RAED} in the Case of Multiple Channels}\label{sec:RAED}
This section considers \textit{RAED} in the case of multiple channels (i.e., \textit{RAED-MCSB} and \textit{RAED-MCMB}). The solution in this case is based on a three-steps procedure inspired by the algorithms developed for the single channel case. In the first step, the channel allocation solution is obtained by a greedy approach. In the second step, the user scheduling for a single EBS is derived by modifying \texttt{ALG-SCSB$_1$}. Finally, the third step is to find the association solution, which is obtained by iteratively applying the two first steps. In order to detail this three-steps solution, we first consider \textit{RAED-MCSB} and describe a heuristic algorithm by modifying \texttt{ALG-SCSB$_1$}. Next, we solve \textit{RAED-MCMB} by proposing an iterative algorithm. 

\subsection{\textit{RAED-MCSB}}
To solve \textit{RAED-MCSB}, we propose a heuristic algorithm, called \texttt{ALG-MCSB}, that first allocates the channels to the users that require the least number of slots. In other words, user $u$ is allocated the channel $c_u$ such that \begin{align}\label{ca}
	c_u=\argmin_{c\in\X{C}}\nu_{u}^c.
\end{align} 
In the case of a tie, user $u$ is allocated the channel that is least used. Once the channel allocation is obtained, \texttt{ALG-MCSB} schedules the users to the EBS by applying a modified version of \texttt{ALG-SCSB$_1$}. The schedule $\Sigma$ is now a matrix $[\sigma_{tc}]$ where $\sigma_{tc}=u$ iff user $u$ is scheduled at slot $t$ using channel $c$. Hence, \texttt{ALG-MCSB} (including the rescheduling procedure) works with the matrix $[\sigma_{tc}]$ and has to verify that two users cannot be scheduled at the same slot using the same channel. 

\subsection{\textit{RAED-MCMB}}

This subsection solves \textit{RAED-MCMB} by applying iteratively \texttt{ALG-MCSB} in each EBS. In every iteration, two main steps are preformed: (1) the EBS that maximizes the number of served users is found and (2) the served users are removed from the whole set of users. The idea of this iterative algorithm (called \texttt{ALG-MCMB}) is similar to the one of the previously proposed approximation algorithm \texttt{ALG-SCMB}. \texttt{ALG-MCMB} is described in the following pseudo-code where the same notations as in Algorithm~\ref{alg} are used. 
\begin{algorithm}[ht!]
	\caption{Heuristic algorithm for \textit{RAED-MCMB}}
	\label{alg:raed}
	\begin{algorithmic}[1]
		\Function{\texttt{ALG-MCMB}}{$\X{U}, \X{B}, \X{C}, \pmb{\nu}, \mathbf{A}$}
		\State Set $\X{X}\gets\X{U}$, $\X{Y}\gets\X{B}$, $\X{Z}\gets\emptyset$, and $\X{A}\gets\emptyset$
		\While{$\X{X}\neq\emptyset$ \textbf{and} $\X{Y}\neq\emptyset$}
		\For{$b$ \textbf{in} $\X{Y}$}
		\State $(\pmb{\Sigma},\X{A}_{b})\gets$ \texttt{ALG-MCSB}$(\X{X},\mathbf{d},\mathbf{A}_{b},\pmb{\nu}^b)$\label{line1:alg:raed}
		\EndFor
		\State $b^\star\gets\argmax\left\{\X{A}_{\X{Y}}\right\}$
		\State $\X{Y}\gets\X{Y}\backslash\{b^\star\}$
		\State $\X{X}\gets\X{X}\backslash\X{A}_{b^\star}$
		\State $\X{Z}\gets\X{Z}\cup\{b^\star\}$
		\EndWhile
		\State\Return $\X{A}_{\X{Z}}$
		\EndFunction
	\end{algorithmic}
\end{algorithm}

Note that \texttt{ALG-MCMB} calls \texttt{ALG-MCSB} in line~\ref{line1:alg:raed} where the latter finds the channel allocation according to~\eqref{ca}. Based on the analysis of the worst-case complexity of \texttt{ALG-SCSB$_1$}, we obtain the worst-case complexity of \texttt{ALG-MCSB} as $O(C+U\log U+UT(U+CT))=O(U^2T+UCT^2)$. Therefore, the worst-case complexity of \texttt{ALG-MCMB} is $O(BL(U^2T+UCT^2))$ where $L\coloneq\min\{U,B\}$.

Of course, since \textit{RAED-MCSB} is $\mathscr{NP}$-hard, neither \texttt{ALG-MCSB} nor \texttt{ALG-MCMB} is guaranteed to give the optimal solution to \textit{RAED-MCSB} or \textit{RAED-MCMB}. Nonetheless, note that the greedy strategy used to allocate the channels can be seen as the best possible locally, i.e., for user $u$, it is optimal to choose the channel that uses fewer number of slots in order to satisfy its request.

In the next subsection, we illustrate how the scheduling solutions obtained in the previous parts of the paper can be modified to be non-preemptive.
\subsection{Non-Preemptive Scheduling}\label{sec:SAED:nonp}
In this section, we discuss \textit{RAED} with non-preemptive scheduling constraints. In other words, we further constrain the problem by imposing that users' requests should not be interrupted. Without loss of generality, we consider the case of $B=C=1$. We start by the following definition.
\begin{definition}[Starting and completion times]~\newline
	\indent Given a schedule $\Sigma=[\sigma_1,\sigma_2,\ldots,\sigma_T]$, we define the starting time $S_u$ (resp. completion time $C_u$) of user $u$ as the smallest (resp. the largest) time for which $\sigma_{S_u}=u$ (resp. $\sigma_{C_u}=u$).
\end{definition}

We show that the schedule returned by \texttt{ALG-SCSB$_1$} can be modified, in polytime, to obtain a new non-preemptive schedule without discarding any scheduled user. Let us first consider \texttt{ALG-SCSB$_1$}. Remember that $\Sigma^u$ is the schedule that corresponds to the set of scheduled users $\X{S}^u$ given by \texttt{ALG-SCSB$_1$} at the end of iteration $u$.

\begin{lemma}\label{lemma:p2nonp}
	The schedule $\Sigma^u$ can be transformed, in polytime, to a non-preemptive one without discarding any user.
\end{lemma}

\begin{IEEEproof}\label{proof:lemma:p2nonp}
	Assume without loss of generality that an arbitrary user $i$ is scheduled preemptively in $\Sigma^u$. Let $S_i$ and $C_i$ be the starting and completion times of $i$. Since the scheduling of $i$ is preempted, then $C_i-S_i+1>\nu_i$ and $C_i-S_i-\nu_i+1$ are the number of slots that remain empty in $\{S_i,\ldots,C_i\}$ due to preemption. First, note that no user other than $i$ is scheduled in the idle slots of $\{S_i,\ldots,C_i\}$ because, otherwise, $i$ would have been scheduled instead. Second, $i$ can be scheduled at time $S_i'=C_i-\nu_i+1$. Doing so, we guarantee that: (i) $i$ is scheduled non-preemptively since $C_i-S_i'+1=\nu_i$, (ii) the deadline constraint of $i$ is always met since $C_i$ is kept unchanged, and (iii) the energy constraint is also met since the slots used in $\{S_i,\ldots,S_i'\}$ can be used in $\{S_i',\ldots,C_i\}$. This proves that the schedule $\Sigma^u$ can be transformed into a non-preemptive one in polytime without discarding any scheduled user. Hence, the lemma is proved.
\end{IEEEproof}

Lemma~\ref{lemma:p2nonp} shows that even if a scheduling solution to \textit{RAED} is preempted, it is possible to change it to a non-preemptive one without decreasing the objective function. This is due primarily to the structure of the proposed algorithms and to the sufficiently large battery capacity.

In the next section, we present simulation results to show the performance of the proposed algorithm.

\section{Simulation Results}\label{sec:sim}
\begin{table}[hbt!]
	\centering
	\begin{tabularx}{\columnwidth}{l|l}
		Parameter & Value \\
		\hline
		EBS transmit power & $30$ dBm \\ \hline
		Total bandwidth $W$ & $20$ MHz \\ \hline
		Carrier frequency & $2$ GHz \\ \hline
		Noise power density & $-174$ dBm/Hz \\ \hline
		Path loss between user $u$ and EBS $b$ & $30.6 + 36.7\log_{10}dist_{u,b}$\\ \hline
		Energy arrival rate & $\lambda=0.5$\\\hline
		Request data size & $s_u\sim\mathrm{unif}\{1Kb, 1Mb\}$ for all $u\in\X{U}$\\\hline    
		Frame length & $T=10$\\\hline
		Deadlines & $d_u\sim\mathrm{unif}\{1,T\}$ for all $u\in\X{U}$\\\hline
		Number of EBSs & $B=10$\\\hline
	\end{tabularx}
	\vskip1em
	\caption{Simulation parameters.}
	\label{table:parameters}
\end{table}

In this section, we present simulation results to illustrate the performance of the proposed algorithms. We consider a geographical area of size $20\times20$ square meters where the users and EBSs are randomly and uniformly distributed. The channel gains are based on 3GPP specifications~\cite{3GPP} and are modeled similarly to~\cite{Zhuang:34:4}. The energy arrival $E_{b,t}$ is assumed to follow a Poisson distribution with parameter $\lambda$. Unless otherwise specified, simulations use the parameters shown in Table~\ref{table:parameters}. The optimization problem~\eqref{pb:2} is modeled in Python using PuLP~\cite{pulp} and solved using the CPLEX solver~\cite{cplex}. The simulations are performed on $1000$ realizations and averaged out, where in each realization independent random parameters are generated.

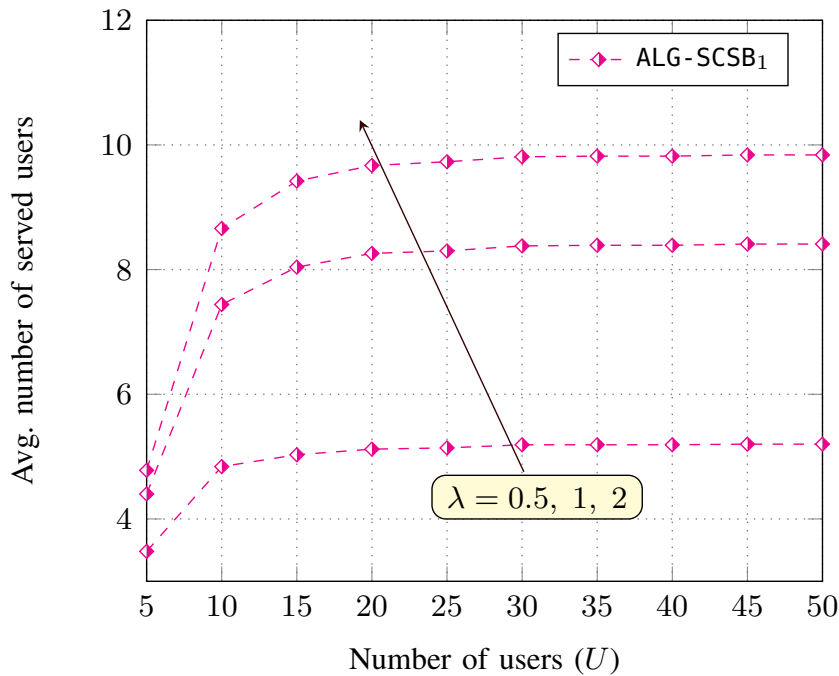
\begin{figure}[H]
  \centering
  \resizebox{.65\textwidth}{!}{%
    \begin{tikzpicture}
      \tikzset{every pin/.style={draw,fill=yellow!20,rectangle,rounded corners=3pt,font=\scriptsize},} 
      \begin{axis}[
        xlabel={Number of users ($U$)},
        ylabel={Avg. number of served users},
        set layers,
        grid=both,
        legend cell align=left,
        xmin=5,
        xmax=50,
        ymin=3,
        ymax=12,
        xtick={5,10,...,50},
        x label style={font=\footnotesize},
        y label style={font=\footnotesize}, 
        ticklabel style={font=\footnotesize},
        legend style={at={(.95,.93)},anchor=east,font=\scriptsize},
        ]
        \addplot[dashed,color=magenta,mark=halfsquare right*,mark options={scale=1,solid}] coordinates {
          (5,3.48) (10,4.84) (15,5.03) (20,5.12) (25,5.14) (30,5.19) (35,5.19) (40,5.19) (45,5.2) (50,5.2)
        };
        \addlegendentry{\texttt{ALG-SCSB$_1$}}
        \addplot[dashed,color=magenta,mark=halfsquare right*,mark options={scale=1,solid}] coordinates {
          (5,4.4) (10,7.44) (15,8.04) (20,8.26) (25,8.3) (30,8.38) (35,8.39) (40,8.39) (45,8.41) (50,8.41)
        };
        \addplot[dashed,color=magenta,mark=halfsquare right*,mark options={scale=1,solid}] coordinates {
          (5,4.78) (10,8.66) (15,9.42) (20,9.67) (25,9.73) (30,9.81) (35,9.82) (40,9.82) (45,9.84) (50,9.84)
        };
        \coordinate (P) at (axis cs:19,10.5);
        \draw [fill=yellow!20,rounded corners] (axis cs:24,4) rectangle (axis cs:38,4.7) node[] {};
        \node (label) at (axis cs:31,4.3) {\footnotesize $\lambda=0.5,\; 1,\; 2$};
        \draw [>=stealth,red!20!black,->,shorten >=2pt] (label) -- (P);
      \end{axis}
    \end{tikzpicture}
  }
  \caption{Average number of served users by \texttt{ALG-SCSB$_1$} for different $\lambda$.}
  \label{fig:scase:lambda}
\end{figure}
Fig.~\ref{fig:scase:lambda} illustrates the performance of \texttt{ALG-SCSB$_1$} when varying both $U$ and $\lambda$. As $\lambda$ increases, more users are served. Since the number of slots is fixed to $T=10$, no more than $10$ users can be served in the best case (when every user requires one unit of energy). It is also clear that as $\lambda$ increases, the number of served users increases rapidly to reach $10$. For example, for a fixed $U$, the gap in the y-axis from $\lambda=0.5$ to $\lambda=1$ is larger than the gap from $\lambda=1$ to $\lambda=2$. This is because as $\lambda$ increases, the energy will be always available and the problem becomes equivalent to scheduling users in increasing order of required energy.

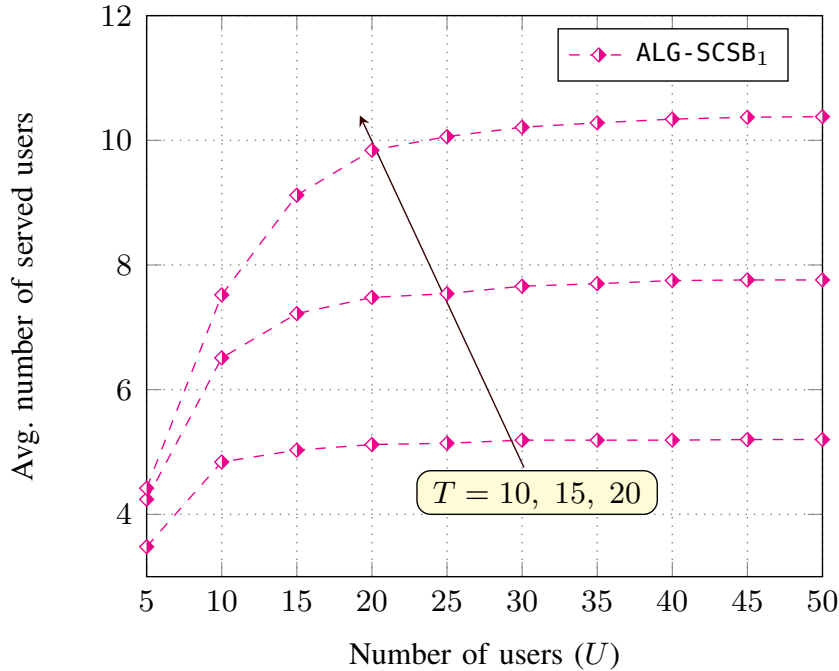
\begin{figure}[htb!]
  \centering
  \resizebox{.65\textwidth}{!}{%
    \begin{tikzpicture}
      \tikzset{every pin/.style={draw,fill=yellow!20,rectangle,rounded corners=3pt,font=\scriptsize},} 
      \begin{axis}[
        xlabel={Number of users ($U$)},
        ylabel={Avg. number of served users},
        set layers,
        grid=both,
        legend cell align=left,
        xmin=5,
        xmax=50,
        ymin=3,
        ymax=12,
        xtick={5,10,...,50},
        x label style={font=\footnotesize},
        y label style={font=\footnotesize}, 
        ticklabel style={font=\footnotesize},
        legend style={at={(.95,.93)},anchor=east,font=\scriptsize},
        ]
        \addplot[dashed,color=magenta,mark=halfsquare right*,mark options={scale=1,solid}] coordinates {
          (5,3.48) (10,4.84) (15,5.03) (20,5.12) (25,5.14) (30,5.19) (35,5.19) (40,5.19) (45,5.2) (50,5.2)
        };
        \addlegendentry{\texttt{ALG-SCSB$_1$}}
        \addplot[dashed,color=magenta,mark=halfsquare right*,mark options={scale=1,solid}] coordinates {
          (5,4.24) (10,6.51) (15,7.22) (20,7.48) (25,7.54) (30,7.66) (35,7.7) (40,7.75) (45,7.76) (50,7.76)
        };
        \addplot[dashed,color=magenta,mark=halfsquare right*,mark options={scale=1,solid}] coordinates {
          (5,4.42) (10,7.52) (15,9.12) (20,9.84) (25,10.06) (30,10.21) (35,10.28) (40,10.34) (45,10.37) (50,10.38)
        };
        \coordinate (P) at (axis cs:19,10.5);
        \draw [fill=yellow!20,rounded corners] (axis cs:23,4) rectangle (axis cs:39,4.7) node[] {};
        \node (label) at (axis cs:31,4.3) {\footnotesize $T=10,\; 15,\; 20$};
        \draw [>=stealth,red!20!black,->,shorten >=2pt] (label) -- (P);
      \end{axis}
    \end{tikzpicture}
  }
  \caption{Average number of served users by \texttt{ALG-SCSB$_1$} for different $T$.}
  \label{fig:scase:T}
\end{figure}
Fig.~\ref{fig:scase:T} shows the performance of \texttt{ALG-SCSB$_1$} against $U$ and $T$. It is clear that as $T$ increases, more users can be served. We can see that, e.g., for $T=20$, the number of served users is about $10$, which indicates that the users require on average $2$ units of energy (i.e., $\nu_u\approx 2$). In fact, as the number of users increases, not all of them can be served even for larger values of $T$. This is due to the required number of slots by each user and also by the energy availability at the EBS.


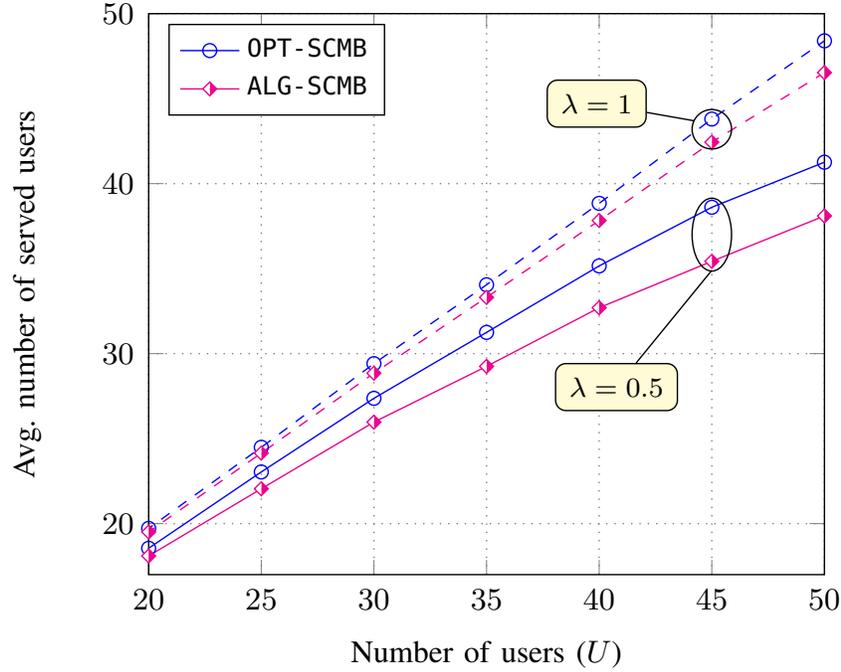
\begin{figure}[t!]
  \centering
  \resizebox{.65\textwidth}{!}{%
    \begin{tikzpicture}
      \tikzset{every pin/.style={draw,fill=yellow!20,rectangle,rounded corners=3pt,font=\scriptsize},} 
      \begin{axis}[
        xlabel={Number of users ($U$)},
        ylabel={Avg. number of served users},
        set layers,
        grid=both,
        legend cell align=left,
        xmin=20,
        xmax=50,
        ymin=17,
        ymax=50,
        xtick={20,25,...,50},
        x label style={font=\footnotesize},
        y label style={font=\footnotesize}, 
        ticklabel style={font=\footnotesize},
        legend style={at={(.35,.90)},anchor=east,font=\scriptsize},
        ]
        \addplot[color=blue,mark=o,mark options={scale=1,solid}] coordinates {
          (5,4.76) (10,9.46) (15,14.18) (20,18.55) (25,23.04) (30,27.37) (35,31.25) (40,35.16) (45,38.61) (50,41.26)
        };
        \addlegendentry{\texttt{OPT-SCMB}}
        \addplot[color=magenta,mark=halfsquare right*,mark options={scale=1,solid}] coordinates {
          (5, 4.73) (10, 9.29) (15, 13.9) (20, 18.1) (25, 22.05) (30, 25.97) (35, 29.25) (40, 32.7) (45, 35.43) (50, 38.1)
        };
        \addlegendentry{\texttt{ALG-SCMB}}
        \addplot[dashed,color=blue,mark=o,mark options={scale=1,solid}] coordinates {
          (5, 4.93) (10, 9.84) (15, 14.77) (20, 19.72) (25, 24.49) (30, 29.41) (35, 34.05) (40, 38.84) (45, 43.8) (50, 48.4)
        };
        \addplot[dashed,color=magenta,mark=halfsquare right*,mark options={scale=1,solid}] coordinates {
          (5, 4.93) (10, 9.79) (15, 14.66) (20, 19.51) (25, 24.15) (30, 28.85) (35, 33.31) (40, 37.83) (45, 42.44) (50, 46.53)
        };
        \node[coordinate] (A) at (axis cs:45,37) {}; 
        \node[coordinate,pin={[align=left,pin distance=1cm,pin edge={black,thin}]250:{$\lambda=0.5$}}] at (axis cs:45,34.9) {}; 
        \node[coordinate] (B) at (axis cs:45,43.2) {}; 
        \node[coordinate,pin={[align=left,pin distance=1cm,pin edge={black,thin},reset label anchor]170:{$\lambda=1$}}] at (axis
        cs:44.2,43.7) {}; 
      \end{axis}
      \draw[black] (A) ellipse (0.2 and 0.37); 
      \draw[black] (B) ellipse (0.2 and 0.2); 
    \end{tikzpicture}
  }
  \caption{Average number of served users by \texttt{ALG-SCMB} and \texttt{OPT-SCMB} for different $\lambda$.}
  \label{fig:gcase:lambda}
\end{figure}
In Fig.~\ref{fig:gcase:lambda}, we compare the performance of \texttt{ALG-SCMB} to \texttt{OPT-SCMB} when varying $U$ and $\lambda$. We can see that \texttt{ALG-SCMB} performs close to \texttt{OPT-SCMB} even for large values of $U$. As $\lambda$ increases, energy will be always available and hence the problem becomes easier; though it is still $\X{NP}$-hard. Consequently, \texttt{ALG-SCMB} performs closer to \texttt{OPT-SCMB} for large values of $\lambda$ than for small ones. Notice that the average performance of \texttt{ALG-SCMB} is better than the theoretical guarantee given by the approximation ratio of $0.5$ (see lemma~\ref{theorem:1}). This is clear from the figure where \texttt{ALG-SCMB} achieves on average a performance ratio (optimal-to-algorithm) of $0.925$ for $U=50$ and $\lambda=0.5$.

\begin{figure}[t!]
  \centering
  \resizebox{.65\textwidth}{!}{%
    \begin{tikzpicture}
      \tikzset{every pin/.style={draw,fill=yellow!20,rectangle,rounded corners=3pt,font=\scriptsize},} 
      \begin{axis}[
        xlabel={Number of users ($U$)},
        ylabel={Avg. number of served users},
        set layers,
        grid=both,
        legend cell align=left,
        xmin=20,
        xmax=50,
        ymin=17,
        ymax=50,
        xtick={20,25,...,50},
        x label style={font=\footnotesize},
        y label style={font=\footnotesize}, 
        ticklabel style={font=\footnotesize},
        legend style={at={(.35,.9)},anchor=east,font=\scriptsize},
        ]
        \addplot[color=blue,mark=o,mark options={scale=1,solid}] coordinates {
          (5,4.76) (10,9.46) (15,14.18) (20,18.55) (25,23.04) (30,27.37) (35,31.25) (40,35.16) (45,38.61) (50,41.26)
        };
        \addlegendentry{\texttt{OPT-SCMB}}
        \addplot[color=magenta,mark=halfsquare right*,mark options={scale=1,solid}] coordinates {
          (5,4.73) (10,9.29) (15,13.9) (20,18.1) (25,22.05) (30,25.97) (35,29.25) (40,32.7) (45,35.43) (50,38.1)
        };
        \addlegendentry{\texttt{ALG-SCMB}}
        \addplot[dashed,color=blue,mark=o,mark options={scale=1,solid}] coordinates {
          (5,4.85) (10,9.71) (15,14.45) (20,19.26) (25,23.99) (30,28.69) (35,33.43) (40,37.92) (45,42.52) (50,46.76)
        };
        \addplot[dashed,color=magenta,mark=halfsquare right*,mark options={scale=1,solid}] coordinates {
          (5,4.82) (10,9.65) (15,14.36) (20,19.03) (25,23.52) (30,27.97) (35,32.42) (40,36.43) (45,40.64) (50,44.48)
        };
        \node[coordinate] (A) at (axis cs:45,37) {}; 
        \node[coordinate,pin={[align=left,pin distance=1cm,pin edge={black,thin}]250:{$T=10$}}] at (axis cs:45,34.8) {}; 
        \node[coordinate] (B) at (axis cs:45,41.5) {}; 
        \node[coordinate,pin={[align=left,pin distance=1cm,pin edge={black,thin},reset label anchor]170:{$T=15$}}] at (axis
        cs:44.2,42) {}; 
      \end{axis}
      \draw[black] (A) ellipse (0.2 and 0.38); 
      \draw[black] (B) ellipse (0.2 and 0.28); 
    \end{tikzpicture}
  }
  \caption{Average number of served users by \texttt{ALG-SCMB} and \texttt{OPT-SCMB} for different $T$.}
  \label{fig:gcase:T}
\end{figure}
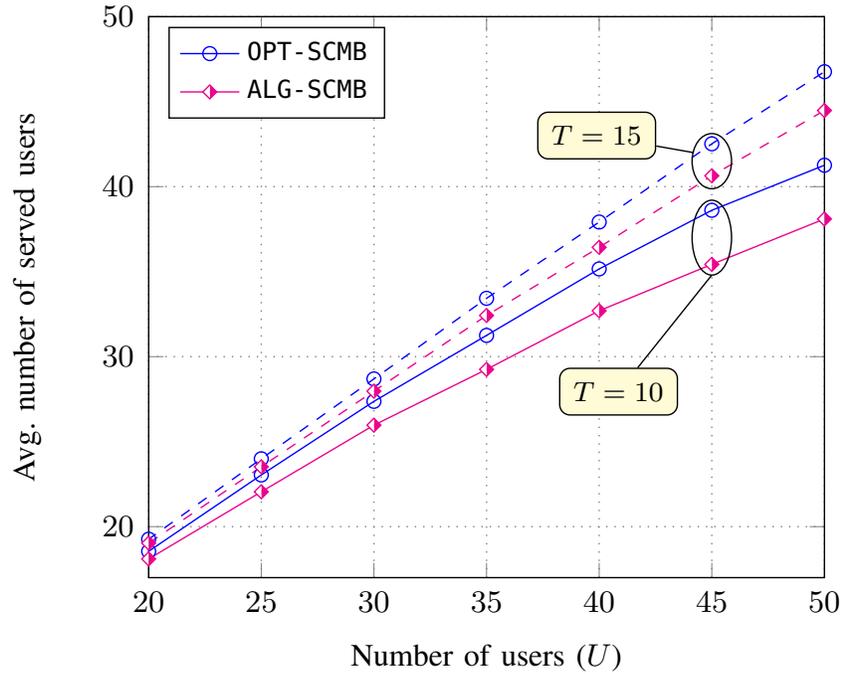
Fig.~\ref{fig:gcase:T} compares \texttt{ALG-SCMB} to \texttt{OPT-SCMB} for different values of $U$ and $T$. Both algorithms are close to each other. When $T$ gets larger and for fixed $U$ and $B$, the gap between both algorithms shrinks. For $T=15$ and despite the small gap between both algorithms, we can see that it increases slowly as $U$ increases. Interestingly, similarly to Fig.~\ref{fig:gcase:lambda}, \texttt{ALG-SCMB} has a better average performance compared to the theoretical performance guarantee.

\begin{figure}[t!]
  \centering
  \resizebox{.65\textwidth}{!}{%
    \begin{tikzpicture}
      \tikzset{every pin/.style={draw,fill=yellow!20,rectangle,rounded corners=3pt,font=\scriptsize},} 
      \begin{axis}[
        xlabel={Number of users ($U$)},
        ylabel={Avg. number of served users},
        set layers,
        grid=both,
        legend cell align=left,
        xmin=20,
        xmax=50,
        ymin=16,
        ymax=47,
        xtick={20,25,...,50},
        x label style={font=\footnotesize},
        y label style={font=\footnotesize}, 
        ticklabel style={font=\footnotesize},
        legend style={at={(.35,.9)},anchor=east,font=\scriptsize},
        ]
        \addplot[color=blue,mark=o,mark options={scale=1,solid}] coordinates {
          (5, 4.76) (10, 9.46) (15, 14.18) (20, 18.55) (25, 23.04) (30, 27.37) (35, 31.25) (40, 35.16) (45, 38.61) (50, 41.26)
        };
        \addlegendentry{\texttt{OPT-SCMB}}
        \addplot[color=magenta,mark=halfsquare right*,mark options={scale=1,solid}] coordinates {
          (5, 4.73) (10, 9.29) (15, 13.9) (20, 18.1) (25, 22.05) (30, 25.97) (35, 29.25) (40, 32.7) (45, 35.43) (50, 38.1)
        };
        \addlegendentry{\texttt{ALG-SCMB}}
        \addplot[densely dotted,color=blue,mark=o,mark options={scale=1,solid}] coordinates {
          (5, 4.77) (10, 9.29) (15, 13.7) (20, 17.29) (25, 20.03) (30, 22.0) (35, 23.27) (40, 23.81) (45, 24.05) (50, 24.23)
        };
        \addplot[densely dotted,color=magenta,mark=halfsquare right*,mark options={scale=1,solid}] coordinates {
          (5, 4.72) (10, 9.1) (15, 13.17) (20, 16.19) (25, 18.78) (30, 20.77) (35, 22.15) (40, 22.93) (45, 23.47) (50, 23.79)
        };
        \addplot[dashed,color=blue,mark=o,mark options={scale=1,solid}] coordinates {
          (5, 4.82) (10, 9.61) (15, 14.21) (20, 18.89) (25, 23.5) (30, 27.93) (35, 32.54) (40, 36.87) (45, 41.47) (50, 45.44)
        };
        \addplot[dashed,color=magenta,mark=halfsquare right*,mark options={scale=1,solid}] coordinates {
          (5, 4.81) (10, 9.57) (15, 14.08) (20, 18.67) (25, 22.89) (30, 27.16) (35, 31.25) (40, 35.05) (45, 39.25) (50, 42.58)
        };
        \node[coordinate] (A) at (axis cs:50,39.6) {}; 
        \node[coordinate,pin={[align=left,pin distance=1cm,pin edge={black,thin}]250:{$B=10$}}] at (axis cs:50,37.5) {}; 
        \node[coordinate] (B) at (axis cs:45,40.35) {}; 
        \node[coordinate,pin={[align=left,pin distance=1cm,pin edge={black,thin},reset label anchor]170:{$B=15$}}] at (axis
        cs:44.2,41) {}; 
        \node[coordinate] (C) at (axis cs:45,23.9) {}; 
        \node[coordinate,pin={[align=left,pin distance=1cm,pin edge={black,thin},reset label anchor]225:{$B=5$}}] at (axis
        cs:44.5,23.1) {}; 
      \end{axis}
      \draw[black] (A) ellipse (0.2 and 0.4); 
      \draw[black] (B) ellipse (0.2 and 0.3); 
      \draw[black] (C) ellipse (0.15 and 0.2); 
    \end{tikzpicture}
  }
  \caption{Average number of served users by \texttt{ALG-SCMB} and \texttt{OPT-SCMB} for different $B$.}
  \label{fig:gcase:N}
\end{figure}
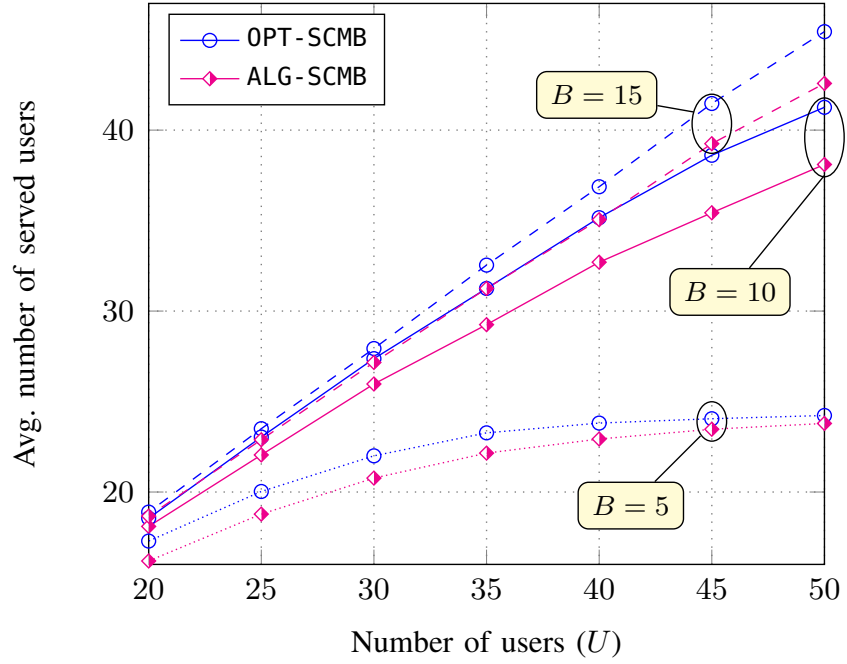
Fig.~\ref{fig:gcase:N} illustrates the performance of \texttt{ALG-SCMB} against \texttt{OPT-SCMB} when varying $U$ and $B$. For small and fixed $B$, \texttt{ALG-SCMB} gets closer to \texttt{OPT-SCMB} as $U$ increases. This is because the maximum possible number of users that can be served is reached and hence the performance of the optimal algorithm saturates. We can see that as more EBSs are available, more users are served by both algorithms and, since $T$ is fixed, \texttt{ALG-SCMB} gets closer to \texttt{OPT-SCMB}. The gap between both algorithms is still small even for large values of $U$ and $B$, which illustrates the superiority of \texttt{ALG-SCMB}.


\begin{figure}[t!]
	\centering
	\resizebox{.65\textwidth}{!}{%
		\begin{tikzpicture}
		\tikzset{every pin/.style={draw=black,fill=yellow!20,rectangle,rounded corners=3pt,font=\scriptsize},} 
		\begin{axis}[
		xlabel={Number of users ($U$)},
		ylabel={Avg. number of served users},
		set layers,
		grid=both,
		legend cell align=left,
		xmin=2,
		xmax=20,
		ymin=1,
		ymax=20,
		xtick={2,4,...,20},
		x label style={font=\footnotesize},
		y label style={font=\footnotesize}, 
		ticklabel style={font=\footnotesize},
		legend style={at={(.35,.9)},anchor=east,font=\scriptsize},
		]
		\addplot[color=blue,mark=o,mark options={scale=1,solid}] coordinates {
			(2, 1.66) (4, 2.92) (6, 3.88) (8, 4.4) (10, 4.78) (12, 5.0) (14, 5.16) (16, 5.32) (18, 5.4) (20, 5.5)
		};
		\addlegendentry{\texttt{OPT-MCMB}}
		\addplot[color=magenta,mark=halfsquare right*,mark options={scale=1,solid}] coordinates {
			(2, 1.64) (4, 2.82) (6, 3.76) (8, 4.14) (10, 4.42) (12, 4.58) (14, 4.68) (16, 4.76) (18, 4.82) (20, 4.84)
		};
		\addlegendentry{\texttt{ALG-MCMB}}
		\addplot[densely dotted,color=blue,mark=o,mark options={scale=1,solid}] coordinates {
			(2, 1.92) (4, 3.8) (6, 5.66) (8, 7.36) (10, 9.22) (12, 11.1) (14, 12.73) (16, 14.1) (18, 15.34) (20, 16.34)
		}; 
		\addplot[densely dotted,color=magenta,mark=halfsquare right*,mark options={scale=1,solid}] coordinates {
			(2, 1.92) (4, 3.8) (6, 5.56) (8, 7.18) (10, 8.92) (12, 10.58) (14, 12.04) (16, 13.14) (18, 14.1) (20, 15.18)
		};
		\node[coordinate] (A) at (axis cs:10,4.6) {}; 
		\node[coordinate,pin={[align=left,pin distance=0.3cm,pin edge={black,thin}]250:{$B=1$}}] at (axis cs:10,3.95) {}; 
		\node[coordinate] (B) at (axis cs:12,10.8) {}; 
		\node[coordinate,pin={[align=left,pin distance=1cm,pin edge={black,thin},reset label anchor]290:{$B=4$}}] at (axis
		cs:12,10.15) {}; 

		\end{axis}
		\draw[black] (A) ellipse (0.1 and 0.2); 
		\draw[black] (B) ellipse (0.1 and 0.2); 
		
		\end{tikzpicture}
	}
	\caption{Average number of served users by \texttt{ALG-MCMB} and \texttt{OPT-MCMB} for different $B$ ($C=2$).}
	\label{fig:raed:B}
\end{figure}
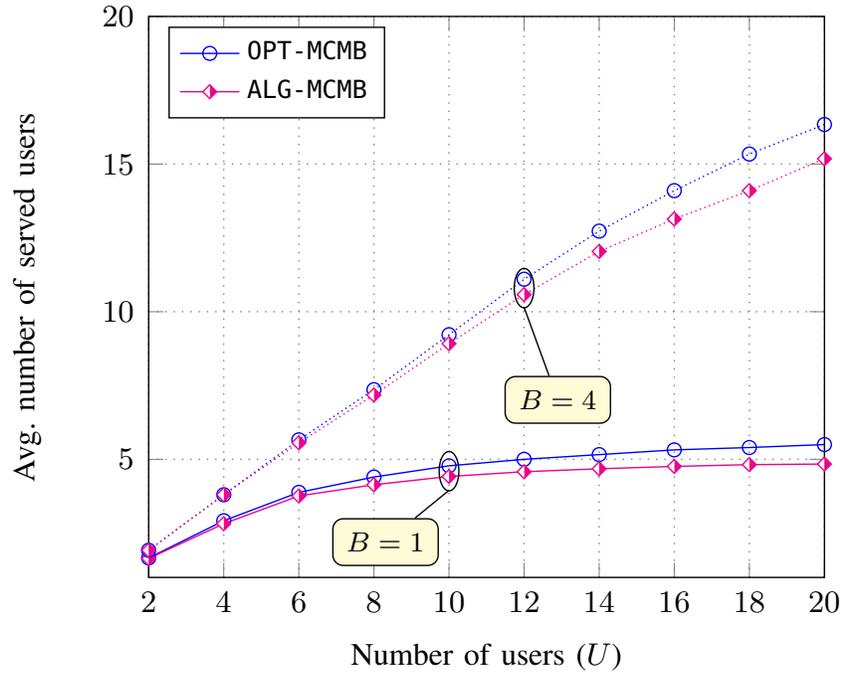
In Fig.~\ref{fig:raed:B}, we illustrate the performance of \texttt{ALG-MCMB} against \texttt{OPT-MCMB}. We can see that the performance of \texttt{ALG-MCMB} is close-to-optimal. When the number of EBSs increases, more users are served even when the energy is constant (i.e., for fixed $\lambda$). From Fig.~\ref{fig:raed:B}, we can see that \texttt{ALG-MCSB} has close-to-optimal performance with a ratio (heuristic-to-optimal) of $0.88$ for $U=20$ and $B=1$. This ratio becomes equal to $0.93$ for $U=20$ and $B=4$. Hence, we conclude that the performance ratio increases with more EBSs in the network, which illustrates the efficiency of \texttt{ALG-MCMB}. 

\begin{figure}[ht!]
	\centering
	\resizebox{.65\textwidth}{!}{%
		\begin{tikzpicture}[every pin/.style={draw,fill=yellow!20,rectangle,rounded corners=3pt,font=\scriptsize},spy using outlines={circle, magnification=9, connect spies}]
		\begin{axis}[
		xlabel={Energy arrival rate ($\lambda$)},
		ylabel={Avg. number of served users},
		axis on top,
		grid=both,
		legend cell align=left,
		xmin=0.5,
		xmax=8,
		ymin=25,
		ymax=140,
		xmode=log,
		log ticks with fixed point,
		xtick={0.5,1,2,4,8},
		x label style={font=\footnotesize},
		y label style={font=\footnotesize}, 
		ticklabel style={font=\footnotesize},
		legend style={at={(.35,.93)},anchor=east,font=\scriptsize},
		]

		\addplot[color=magenta] coordinates {
			(0.5, 28.08) (1, 48.38) (2, 57.78) (4, 59.1) (8, 59.48)
		};
		\addlegendentry{\texttt{ALG-MCMB}}
		\addplot[color=magenta,mark=halfsquare right*,mark options={scale=1,solid}] coordinates {
			(0.5, 29.12) (1, 56.08) (2, 86.06) (4, 92.18) (8, 93.26)
		};
		\addplot[color=magenta,mark=diamond,mark options={scale=1,solid}] coordinates {
			(0.5, 29.22) (1, 58.0) (2, 92.14) (4, 99.26) (8, 99.64)
		};
		
		\addplot[color=magenta,mark=o,mark options={scale=1,solid}] coordinates {
			(0.5, 46.3) (1, 73.98) (2, 85.02) (4, 87.52) (8, 87.61)
		};
		\addplot[color=magenta,mark=asterisk,mark options={scale=1,solid}] coordinates {
			(0.5, 47.66) (1, 81.46) (2, 95.86) (4, 97.94) (8, 98.42)
		};
		\addplot[color=magenta,mark=oplus,mark options={scale=1,solid}] coordinates {
			(0.5, 47.74) (1, 82.36) (2, 97.64) (4, 99.74) (8, 99.98)
		};
		
		\coordinate (P) at (axis cs:1.09,68);
		\draw [fill=yellow!20,rounded corners] (axis cs:2,40.5) rectangle (axis cs:4,50.5) node[] {};
		\node (label) at (axis cs:2.8,45.1) {\footnotesize $C=1,\; 2,\; 10$};
		\draw [>=stealth,red!20!black,->,shorten >=2pt] (label) -- (P);
		\coordinate (P1) at (axis cs:1.1,72);
		\draw [fill=yellow!20,rounded corners] (axis cs:1.2,109) rectangle (axis cs:2.4,119) node[] {};
		\node (label1) at (axis cs:1.7,114.6) {\footnotesize $C=10,\; 2,\; 1$};
		\draw [>=stealth,red!20!black,->,shorten >=2pt] (label1) -- (P1);
		\node[coordinate] (A) at (axis cs:1,53) {}; 
		\node[coordinate,pin={[align=left,pin distance=0.5cm,pin edge={black,thin}]300:{$B=6$}}] at (axis cs:1,46.5) {}; 
		\node[coordinate] (B) at (axis cs:1,78) {}; 
		\node[coordinate,pin={[align=left,pin distance=0.5cm,pin edge={black,thin}]130:{$B=10$}}] at (axis cs:1,84.5) {}; 
		
		\end{axis}
		\draw[black] (A) ellipse (0.2 and 0.32); 
		\draw[black] (B) ellipse (0.2 and 0.32); 
		
		\end{tikzpicture}
	}
	\caption{Average number of served users by \texttt{ALG-MCMB} for different $B$ and $C$ ($U=100$).}
	\label{fig:raed:C}
\end{figure}
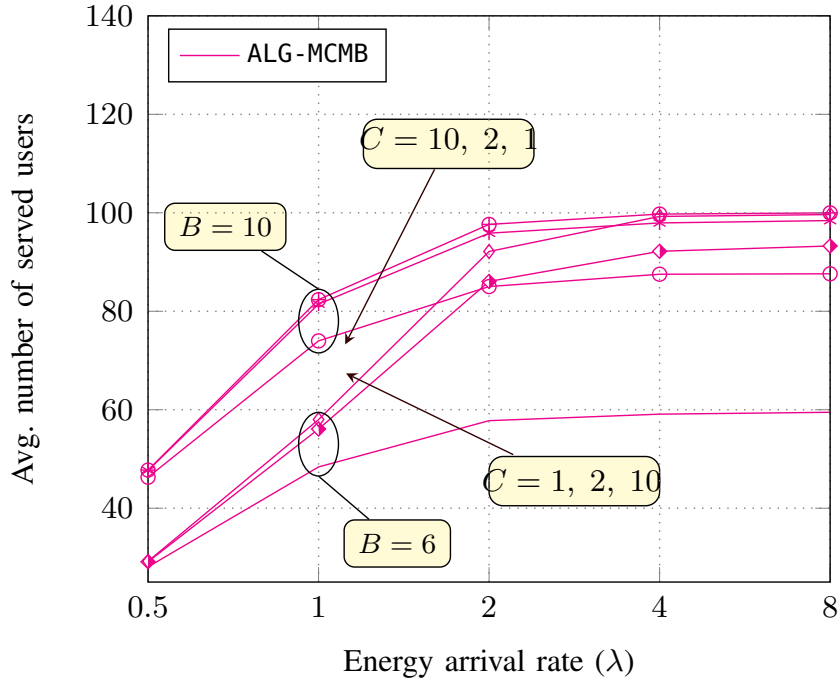
Fig.~\ref{fig:raed:C} illustrates the performance of \texttt{ALG-MCMB} for large number of users, $U=100$. (The optimal algorithm cannot be obtained for computational running-time complexity issues.) We can observe that \texttt{ALG-MCMB} serves more users as $\lambda$, $B$, or $C$ increases. Importantly, the number of served users by \texttt{ALG-MCMB} increases faster with $B$ than with $C$. For example, on the one hand, for $\lambda=1$ and when $B$ varies from $6$ to $10$, \texttt{ALG-MCMB} serves $1.45$ times more users for $C=2$ and $1.42$ times more users for $C=10$. On the other hand, for $\lambda=1$ and when $C$ varies from $2$ to $10$, \texttt{ALG-MCMB} serves $1.035$ times more users for $B=6$ and serves $1.011$ times more users for $B=10$. We can conclude form Fig.~\ref{fig:raed:C} that (i) having more channels is advantageous (in terms of maximizing the number of users) when $\lambda$ is large and $B$ is fixed, and (ii) having more BSs is advantageous when $\lambda$ is small and $C$ is fixed.

\section{Conclusion}\label{sec:cl}
In this paper, we studied a resource allocation problem in green dense cellular networks. This problem, called resource allocation with energy and deadlines constraints (\textit{RAED}), involved user association, scheduling with hard deadlines and channel allocation where base stations (BSs) are solely powered by harvested energy. First, we modeled \textit{RAED} as integer linear program. Next, we characterized the computational complexity of four cases of \textit{RAED} by studying their $\X{NP}$-hardness. For the case of a single channel and a single BS, we proposed two optimal polynomial-time algorithms---the second being much less complex that only works under a common deadlines assumption. These two proposed algorithms were shown to have robust performance against various parameters changes. For the case of a single channel and multiple BSs, \textit{RAED} was shown to be $\X{NP}$-hard and a $\frac{1}{2}$-approximation algorithm was designed to solve it. The average performance of this approximation algorithm was shown to be higher than the theoretical performance guarantee. Next, we studied the case of multiple channels (with both single and multiple BSs). We showed that \textit{RAED} is $\X{NP}$-hard even with a single BS and two channels and we designed a heuristic algorithm to solve both cases. We showed that this heuristic algorithm has close-to-optimal performance. 

As for future work, we will solve \textit{RAED} in the online settings with fairness constraints between users. Our objective will be to develop competitive and fair algorithms that perform well compared to a omniscient offline algorithm.

\bibliographystyle{IEEEtran}
\bibliography{IEEEabrv,ASE}

\begin{thebibliography}{10}
\providecommand{\url}[1]{#1}
\csname url@samestyle\endcsname
\providecommand{\newblock}{\relax}
\providecommand{\bibinfo}[2]{#2}
\providecommand{\BIBentrySTDinterwordspacing}{\spaceskip=0pt\relax}
\providecommand{\BIBentryALTinterwordstretchfactor}{4}
\providecommand{\BIBentryALTinterwordspacing}{\spaceskip=\fontdimen2\font plus
\BIBentryALTinterwordstretchfactor\fontdimen3\font minus
  \fontdimen4\font\relax}
\providecommand{\BIBforeignlanguage}[2]{{%
\expandafter\ifx\csname l@#1\endcsname\relax
\typeout{** WARNING: IEEEtran.bst: No hyphenation pattern has been}%
\typeout{** loaded for the language `#1'. Using the pattern for}%
\typeout{** the default language instead.}%
\else
\language=\csname l@#1\endcsname
\fi
#2}}
\providecommand{\BIBdecl}{\relax}
\BIBdecl

\bibitem{Andrews:32:6}
J.~G. Andrews, S.~Buzzi, W.~Choi, S.~V. Hanly, A.~Lozano, A.~C.~K. Soong, and
  J.~C. Zhang, ``{W}hat {W}ill {5G} {B}e?'' \emph{{IEEE} J. Sel. Areas
  Commun.}, vol.~32, no.~6, pp. 1065--1082, Jun. 2014.

\bibitem{Damnjanovic:18:3}
A.~Damnjanovic, J.~Montojo, Y.~Wei, T.~Ji, T.~Luo, M.~Vajapeyam, T.~Yoo,
  O.~Song, and D.~Malladi, ``{A} {S}urvey {O}n {3GPP} {H}eterogeneous
  {N}etworks,'' \emph{{IEEE} Wireless Commun.}, vol.~18, no.~3, pp. 10--21,
  Jun. 2011.

\bibitem{Liu:35:6}
G.~Liu, X.~Hou, J.~Jin, F.~Wang, Q.~Wang, Y.~Hao, Y.~Huang, X.~Wang, X.~Xiao,
  and A.~Deng, ``{3-D-MIMO With Massive Antennas Paves the Way to 5G Enhanced
  Mobile Broadband: From System Design to Field Trials},'' \emph{{IEEE} J. Sel.
  Areas Commun.}, vol.~35, no.~6, pp. 1222--1233, Jun. 2017.

\bibitem{Elsawy:28:36}
H.~Elsawy, E.~Hossain, and D.~I. Kim, ``{HetNets with Cognitive Small Cells:
  User Offloading and Distributed Channel Access Techniques},'' \emph{{IEEE}
  Commun. Mag.}, vol.~51, no.~6, pp. 28--36, Jun. 2013.

\bibitem{Zhuang:2931:2942}
B.~Zhuang, D.~Guo, E.~Wei, and M.~L. Honig, ``{Scalable Spectrum Allocation and
  User Association in Networks With Many Small Cells},'' \emph{{IEEE} Trans.
  Commun.}, vol.~65, no.~7, pp. 2931--2942, Jul. 2017.

\bibitem{Wang:11:2}
B.~Wang, Q.~Kong, W.~Liu, and L.~T. Yang, ``{On Efficient Utilization of Green
  Energy in Heterogeneous Cellular Networks},'' \emph{IEEE Systems Journal},
  vol.~11, no.~2, pp. 846--857, Jun. 2017.

\bibitem{Zhuang:5470:5483}
B.~Zhuang, D.~Guo, E.~Wei, and M.~L. Honig, ``{Large-Scale Spectrum Allocation
  for Cellular Networks via Sparse Optimization},'' \emph{{IEEE} Trans. Signal
  Process.}, vol.~66, no.~20, pp. 5470--5483, Oct. 2018.

\bibitem{Zhuang:34:4}
B.~Zhuang, D.~Guo, and M.~L. Honig, ``{E}nergy-{E}fficient {C}ell {A}ctivation,
  {U}ser {A}ssociation, and {S}pectrum {A}llocation in {H}eterogeneous
  {N}etworks,'' \emph{{IEEE} J. Sel. Areas Commun.}, vol.~34, no.~4, pp.
  823--831, Apr. 2016.

\bibitem{Zhao:5825:5837}
J.~Zhao, Y.~Liu, K.~K. Chai, A.~Nallanathan, Y.~Chen, and Z.~Han, ``{Spectrum
  Allocation and Power Control for Non-Orthogonal Multiple Access in
  HetNets},'' \emph{{IEEE} Trans. Wireless Commun.}, vol.~16, no.~9, pp.
  5825--5837, Sept. 2017.

\bibitem{Lin:1025:1039}
Y.~Lin, W.~Bao, W.~Yu, and B.~Liang, ``{Optimizing User Association and
  Spectrum Allocation in HetNets: A Utility Perspective},'' \emph{{IEEE} J.
  Sel. Areas Commun.}, vol.~33, no.~6, pp. 1025--1039, Jun. 2015.

\bibitem{Huang:1235:1249}
C.~Huang, J.~Zhang, H.~V. Poor, and S.~Cui, ``{Delay-Energy Tradeoff in
  Multicast Scheduling for Green Cellular Systems},'' \emph{{IEEE} J. Sel.
  Areas Commun.}, vol.~34, no.~5, pp. 1235--1249, May 2016.

\bibitem{Deshmukh:3661:3674}
A.~Deshmukh and R.~Vaze, ``{Online Energy-Efficient Packet Scheduling for a
  Common Deadline With and Without Energy Harvesting},'' \emph{{IEEE} J. Sel.
  Areas Commun.}, vol.~34, no.~12, pp. 3661--3674, Dec. 2016.

\bibitem{Krishnasamy:307:314}
S.~Krishnasamy and S.~Shakkottai, ``{Spectrum Sharing and Scheduling in
  D2D-enabled Dense Cellular Networks},'' in \emph{Proc. Int. Symp. on Modeling
  and Optim. in Mobile, Ad Hoc, and Wireless Netw. (WiOpt)}, May 2015, pp.
  307--314.

\bibitem{Wang:6:5}
Y.~C. Wang and S.~T. Chen, ``{Delay-Aware ABS Adjustment to Support QoS for
  Real-Time Traffic in LTE-A HetNet},'' \emph{{IEEE} Wireless Commun. Lett.},
  vol.~6, no.~5, pp. 590--593, Oct. 2017.

\bibitem{Sciancalepore:193:201}
V.~Sciancalepore, I.~Filippini, V.~Mancuso, A.~Capone, and A.~Banchs, ``{A
  Semi-Distributed Mechanism for Inter-Cell Interference Coordination
  Exploiting the ABSF Paradigm},'' in \emph{Proc. IEEE Int. Conf. on Sensing,
  Commun., and Networking (SECON)}, Jun. 2015, pp. 193--201.

\bibitem{Mlika:21:12}
Z.~Mlika, E.~Driouch, and W.~Ajib, ``{User Association and Scheduling With Hard
  Deadlines in Heterogeneous Cellular Networks},'' \emph{{IEEE} Commun. Lett.},
  vol.~21, no.~12, pp. 2698--2701, Dec. 2017.

\bibitem{Chetto:2:2}
M.~Chetto, ``{Optimal Scheduling for Real-Time Jobs in Energy Harvesting
  Computing Systems},'' \emph{{IEEE} Trans. Emerg. Topics Comput.}, vol.~2,
  no.~2, pp. 122--133, Jun. 2014.

\bibitem{Wang:4:3}
H.~Wang, J.~X. Zhang, and F.~Li, ``{Worst-case Performance Guarantees of
  Scheduling Algorithms Maximizing Weighted Throughput in Energy Harvesting
  Networks},'' \emph{Sustainable Computing: Informatics and Systems}, vol.~4,
  no.~3, pp. 172 -- 182, 2014.

\bibitem{Shan:33:3}
F.~Shan, J.~Luo, W.~Wu, M.~Li, and X.~Shen, ``{Discrete Rate Scheduling for
  Packets With Individual Deadlines in Energy Harvesting Systems},''
  \emph{{IEEE} J. Sel. Areas Commun.}, vol.~33, no.~3, pp. 438--451, Mar. 2015.

\bibitem{Hentati}
A.~Hentati, J.~F. Frigon, , and W.~Ajib, ``{Information Age and Packet Loss
  Performance Analysis of Energy Harvesting WSNs},'' in \emph{Proc. IEEE
  Vehicular Technology Conference (VTC-Fall)}, Sept. 2018, pp. 1--5.

\bibitem{cplex}
\BIBentryALTinterwordspacing
IBM, ``{IBM ILOG CPLEX Optimizer},'' 2010. [Online]. Available:
  \url{http://www-01.ibm.com/software/integration/optimization/cplex-optimizer/}
\BIBentrySTDinterwordspacing

\bibitem{Schrijver:1986:TLI:17634}
A.~Schrijver, \emph{Theory of Linear and Integer Programming}.\hskip 1em plus
  0.5em minus 0.4em\relax New York, NY, USA: John Wiley \& Sons, Inc., 1986.

\bibitem{Chekuri:713:728}
C.~Chekuri and S.~Khanna, ``{A Polynomial Time Approximation Scheme for the
  Multiple Knapsack Problem},'' \emph{SIAM Journal on Computing}, vol.~35,
  no.~3, pp. 713--728, 2005.

\bibitem{Garey:1979}
M.~R. Garey and D.~S. Johnson, \emph{Computers and Intractability: A Guide to
  the Theory of NP-Completeness}.\hskip 1em plus 0.5em minus 0.4em\relax New
  York, NY, USA: W. H. Freeman \& Co., 1979.

\bibitem{Pinedo:2016}
M.~L. Pinedo, \emph{Scheduling: Theory, Algorithms, and Systems}, 5th~ed.\hskip
  1em plus 0.5em minus 0.4em\relax Springer Publishing Company, Incorporated,
  2016.

\bibitem{Williamson:2011:DAA:1971947}
D.~P. Williamson and D.~B. Shmoys, \emph{The Design of Approximation
  Algorithms}, 1st~ed.\hskip 1em plus 0.5em minus 0.4em\relax New York, NY,
  USA: Cambridge University Press, 2011.

\bibitem{3GPP}
3GPP, ``{F}urther {A}dvancements for {E-UTRA} {P}hysical {L}ayer {A}spects
  ({R}elease 9),'' Tech. Rep. 3GPP TR 36.814, Mar. 2010, v9.0.0.

\bibitem{pulp}
S.~Mitchell, M.~O'Sullivan, and I.~Dunning, ``{PuLP: A Linear Programming
  Toolkit for Python},'' The University of Auckland, Department of Engineering
  Science, Tech. Rep., Sept. 2011.

\end{thebibliography}

\end{document}